\documentclass[pra,letterpaper,aps,10pt,superscriptaddress,floatfix,showpacs]{revtex4-1}

\usepackage{graphicx}
\usepackage{amsmath}
\usepackage{amsfonts}
\usepackage{amssymb}
\usepackage{epsfig}
\usepackage[pdftex]{color}
\usepackage{amsmath,graphicx,amssymb,braket,subfigure}
\usepackage{url}

\newcommand{\fref}[1]{Fig. \ref{#1}}

\def\H{\hat{H}}

\def\Heffnh{{\hat{H}^{\rm nh}_{\rm eff}}}
\def\Im{{\rm Im}}

\def\Jj{{\hat{J}_j}}

\def\Jdj{{\hat{J}^\dagger_j}}
\def\LL{\hat{\bar{\mathcal{L}}}}
\def\R{R}
\def\Re{{\rm Re}}

\def\sig{\hat{\sigma}}

\def\siggg{ \sig_{11}}
\def\sigee{ \sig_{22}}
\def\sigge{ \sig_{12}}
\def\sigeg{ \sig_{21}}

\def\sigii{ \sig_{33}}
\def\sigig{ \sig_{31}}
\def\siggi{ \sig_{13}}
\def\sigei{ \sig_{23}}
\def\sigie{ \sig_{32}}

\begin{document}
\title{Continuous multi-step pumping of the optical clock transition in alkaline-earth atoms with minimal perturbation}
\author{Christoph Hotter}
\affiliation{Institut f\"ur Theoretische Physik, Universit\"at Innsbruck, Technikerstr. 21a, A-6020 Innsbruck, Austria}
\author{David Plankensteiner}
\affiliation{Institut f\"ur Theoretische Physik, Universit\"at Innsbruck, Technikerstr. 21a, A-6020 Innsbruck, Austria}
\author{Georgy Kazakov}
\affiliation{Atominstitut, TU Wien, Stadionallee 2, 1020 Vienna, Austria}
\author{Helmut Ritsch}
\affiliation{Institut f\"ur Theoretische Physik, Universit\"at Innsbruck, Technikerstr. 21a, A-6020 Innsbruck, Austria}
\date{\today}

\begin{abstract}
A suitable scheme to continuously create inversion on an optical clock transition with negligible perturbation is a key missing ingredient required to build an active optical atomic clock. Repumping of the atoms on the narrow transition typically needs several pump lasers in a multi step process involving several auxiliary levels. In general this creates large effective level shifts and a line broadening, strongly limiting clock accuracy. Here we present an extensive theoretical study for a realistic multi-level implementation in search of parameter regimes where a sufficient inversion can be achieved with minimal perturbations. Fortunately we are able to identify a useful operating regime, where the frequency shifts remain small and controllable, only weakly perturbing the clock transition for useful pumping rates. For practical estimates of the corresponding clock performance we introduce a straightforward mapping of the multilevel pump scheme to an effective energy shift and broadening parameters for the reduced two-level laser model system. This allows to evaluate the resulting laser power and spectrum using well-known methods.
\end{abstract}

\maketitle

\section{Introduction}
State of the art optical atomic lattice clocks achieve an excellent fractional stability of up to $6.6 \times 10^{-19}$ after one hour of averaging \cite{Oelker2019}. In a typical atomic clock a stable local laser oscillator is compared to the reference transition frequency of trapped ultra-cold atoms. Technically the local laser oscillator is stabilized by an ultra-stable macroscopic cavity with a very good short time stability. Limitations of its stability originate from length fluctuations due to environmental and thermal perturbations \cite{numata2004thermal}. Currently, these perturbations are the central limiting factor of the performance of passive atomic clocks on the short timescale.
It has been proposed \cite{Chen2009active, meiser2009prospects} that active optical clocks, realized as so called superradiant lasers 
\cite{bohnet2012steady, bohnet2014linear, bychek2021superradiant, Chen2009active, debnath2018lasing, Gogyan2020characterisation, Haake1993superradiant, hotter2019superradiant, Jaeger21, kazakov2013active, kazakov2014active, kazakov2016synchro, kazakov2017ions, laske2019pulse, liu2020rugged, maier2014superradiant, meiser2009prospects, meiser2010intensity, meiser2010steady, norcia2016cold, norcia2016superradiance, norcia2018cavity, norcia2018frequency, schaffer2020lasing, tang2021cavity, shankar2021subradiant, weiner2017phase, xu2014synchronization, zhang2021ultranarrow}, can overcome this limitation. In such a laser an ensemble of atoms with a narrow and stable transition is used as gain medium inside an optical resonator. Since the cavity bandwidth is much broader than the gain profile, the frequency of such a bad-cavity laser is primarily determined by the stability of the resonance frequency of the gain medium which makes the system robust against cavity length fluctuations. 

Maintaining population inversion on the atomic clock transition is, of course, a necessary ingredient required for continuous operation of an active optical clock laser. One possibility to achieve this is to prepare the atoms in the upper lasing state outside of the active lasing region and subsequently injecting them into the cavity. Such an approach is reminiscent of the hydrogen maser \cite{goldenberg1960atomic, strelnitski1995hydrogen}. In the optical regime this can be realized, for example, as an atomic beam laser \cite{salzburger2007atom, Chen2009active, liu2020rugged, chen2019continuous}, where atoms in the upper lasing state traverse the cavity.

The most common approach to maintain inversion is to continuously repump the laser active atoms trapped within a magic wavelength optical lattice inside the cavity\cite{meiser2009prospects, meiser2010steady, bohnet2012steady, maier2014superradiant, bohnet2014linear, katori2003ultrastable, ye2008quantum}. Hence each atom can emit several photons into the cavity and for a continuous operation, only a relatively small flux of atoms is needed to compensate the lost atoms. This could be e.g. achieved with an optical conveyor lattice through the cavity\cite{escudero2021steady, bennetts2017steady, chen2019continuous, kazakov2013active, kazakov2014active}. The central challenge of this approach are the perturbations of the clock atoms due to the presence of the repumping lasers.
Typical theoretical models dedicated to superradiant clock lasers with continuous repumping simply assume an artificial transition rate modelled as inverse spontaneous decay from the ground to the excited lasing state \cite{meiser2009prospects, maier2014superradiant, hotter2019superradiant, meiser2010steady, meiser2010intensity}. This introduces an effective homogeneous broadening of the laser line but ignores all shifts and inhomogeneous broadening.  To model this in a more realistic schemes, however, one needs to introduce laser-induced transitions to some auxiliary intermediate levels followed by a spontaneous decay to the upper lasing state. Naturally these lasers will introduce differential light shifts in addition to decoherence on the clock transition. For a non-uniform pump laser field distribution one, of course, gets inhomogeneous broadening. Besides broadening and shifting the laser line, it will eventually modify the threshold and even inhibit lasing.
Frequency shifts lead to additional inaccuracy at least if they are not controllable and precisely measurable, as the resulting laser frequency then differs from the bare atomic transition. Luckily, as long as the inhomogeneous broadening is sufficiently small and symmetric, the atoms with different energies still synchronize \cite{xu2014synchronization, kazakov2016synchro, bychek2021superradiant} and a narrow laser linewidth can be maintained. Therefore, the design of high-performance active optical clocks with a continuous repumping scheme requires the characterization, control and, if possible, minimization of the induced shifts and resulting decoherence.

The conceptually simplest realistic repumping scheme, the so-called three-level scheme, includes only a single intermediate level coherently coupled to the laser ground state, which ideally directly decays to the upper lasing level on a short time scale. Its theoretical study can be drastically simplified when the auxiliary level can be adiabatically eliminated, reducing the model to an effective two-level system subject to an effective incoherent pump, as e.g. in $\rm Lu^+$ ions \cite{kazakov2017ions}. The Stark shift of the lasing transition can be effectively added to the model. However, in most metrology-relevant neutral atoms, as e.g. strontium or ytterbium, such an ideal intermediate level does not exist and thus any realistic repumping scheme requires at least two laser-induced transition steps to irreversibly excite the atoms from the lower to the upper clock state. This has the advantage that one has more possibilities to obtain a desired pump rate with minimal perturbation. However, it has also the disadvantage that, on the one hand, an analytic procedure of adiabatic elimination is cumbersome, especially for systems with a complex multilevel structure. On the other hand, a full numerical treatment, including all the relevant levels in the laser model, significantly increases the computational cost. Especially in time-domain simulations the characteristic time constants of the laser active and intermediate states often differ by many orders of magnitude, requiring a large number of time steps to be calculated. 

Of course the numerical challenges become even more prominent for calculations beyond the mean-field approximation as needed for reliable predictions of linewidth and stability. As particularly useful models to tackle this, we will employ higher order cumulant expansion methods\cite{Kubo1962generalized, plankensteiner2021quantumcumulantsjl}. Luckily we see that, as in the three level case, adiabatic elimination of the intermediate levels can reduce these multi-level systems to a simplified effective two-level system with sufficient accuracy.

In this paper we consider a quite general multi-level repumping scheme for neutral ${\rm ^{88}Sr}$. We demonstrate that a proper choice of intensities and detunings of the pump lasers can lead to a sufficiently high effective repumping rate while at the same time frequency shifts and decoherence rates are kept small. To perform this analysis, we introduce a numerical method to reduce the complex multi-level system to an effective two-level system. This work is organized as follows: In Section~\ref{sec:adiabatic} we review the simplified two-level model including incoherent repumping, and describe the generalized method used to reduce a multilevel system to a two-level one. In Section~\ref{sec:repump_model} we introduce a repumping scheme for trapped ${\rm^{88}Sr}$ and calculate the effective parameters of the equivalent two-level system. The pumping scheme includes the two lasing states and four intermediate states. In Section~{\ref{sec:laser}} we compare the effective two-level laser model with the full six-level laser model.

\section{A multistep excitation process as effective two-level system}
\label{sec:adiabatic}

In this section we describe the method to eliminate the intermediate states in a multilevel scheme with continuous repumping to an effective two-level system by using the eigenvalues of the non-hermitian Hamiltonian. The requirement on this procedure is that the intermediates states can be adiabatically eliminated. The motivation for this procedure is that the full laser system can be numerically very extensive for multilevel systems, and a "conventional" adiabatic elimination is often too cumbersome to be handled analytically. With our method we numerically calculate first the appropriate parameters of an equivalent two-level atom, to use them afterwards in an effective laser model. This has the additional advantage that this simplified model has already been studied extensively \cite{meiser2009prospects, meiser2010intensity, meiser2010steady, bohnet2014linear, bychek2021superradiant}. 

To establish the correspondence between the effective two-level and the multilevel system, we investigate first a two-level atom subjected to spontaneous decay, decoherence and incoherent pumping. In the Heisenberg representation the averaged value of an operator $\hat{O}$ for an open quantum system follows the equation
\begin{equation}
\frac{d\langle \hat{O}\rangle}{dt}=\frac{i}{\hbar} \langle[\H,\hat{O}]\rangle+\langle \LL[\hat{O}]\rangle
\label{e:HeisLangAvEqGeneral}
\end{equation}
\noindent where $\H$ is the Hamiltonian and $\LL$ is the super-operator describing the dissipative processes. Within the Born-Markov approximation $\LL$ has the form
\begin{equation}
\LL[\hat{O}] = \sum_j \frac{R_j}{2} \left(2 \Jdj \hat{O} \Jj-\Jdj \Jj \hat{O} - \hat{O} \Jdj \Jj \right),
\label{e:LindbladGeneral}
\end{equation}
here $\Jj$ are the jump operators with the corresponding rates $R_j$. 
For our two-level atom in the rotating frame of the unperturbed atomic transition frequency the Hamiltonian can be written as $\H=\hbar (\delta_{1} \sig_{11} + \delta_{2} \sig_{22})$, where $\delta_{1}$ and $\delta_2$ are the shifts from the ground $\ket{1}$ and excited clock state $\ket{2}$, respectively and $\sig_{ij} = \ket{i} \bra{j}$. The jump operators and corresponding rates of the dissipative processes are listed in Table~\ref{tab:2level}.
\begin{table}[b]
\begin{center}
\begin{tabular}{ |c|c|c|c| } 
 \hline
 \# & jump & rate & description \\ 
 \hline
 1 & $\sig_{12}$ & $\Gamma_{12}$ & decay from $\ket{2}$ to $\ket{1}$ \\ 
 2 & $\sig_{21}$ & $\R$ & incoherent pumping from $\ket{1}$ to $\ket{2}$ \\
 3 & $\sig_{11}$ & $\nu_{1}$ & dephasing on $\ket{1}$ \\
 4 & $\sig_{22}$ & $\nu_{2}$ & dephasing on $\ket{2}$ \\
\hline
\end{tabular}
\end{center}
\caption{\emph{Dissipative processes of the two-level scheme.} }
\label{tab:2level}
\end{table}
The equations of motion for the operator averages $\braket{\sig_{ij}}$ of such a two-level atom are
\begin{align}
\partial_t \braket{\sig_{22}} & = \R \braket{\sig_{11}} - \Gamma_{12} \braket{\sig_{22}} 
\label{e:TwoLevelSig22} \\
\partial_t \braket{\sig_{12}} & = -\left(\frac{\R + \Gamma_{12} + \nu}{2} +i \delta_{21} \right) \braket{\sig_{12}} ,
\label{e:TwoLevelSig12}
\end{align}
where $\delta_{21} = \delta_2 - \delta_1$, and $\nu = \nu_1+\nu_2$. From (\ref{e:TwoLevelSig22}) one can easily express the incoherent repumping rate $\R$ via the ratio of the steady-state population as
\begin{equation}
R = \frac{\braket{\sig_{22}}}{\braket{\sig_{11}}}\Gamma_{12}.
\label{e:R}
\end{equation}
To express the dephasing rates $\nu_1$ and $\nu_2$ we exploit the effective non-hermitian Hamiltonian 
\begin{equation}
\Heffnh = \H - \frac{i \hbar}{2} \sum_j R_j \Jdj \Jj ,
\label{e:effHgeneral}
\end{equation}
as it is used e.g. in the Monte-Carlo wave function approach \cite{Dum1992montecarlo, Molmer1993MonteCarlo, plenio1998quantumjumps}. 
For our two-level system this non-hermitian Hamiltonian has the form
\begin{equation}
\Heffnh = \hbar \delta_{1} \sig_{11} + \hbar \delta_{2} \sig_{22} - \frac{i \hbar}{2} 
 \left[ \Gamma_{12} \sig_{22} + \R \sig_{11} + \nu_1 \sig_{11} + \nu_2 \sig_{22}) \right],
\label{e:effH2level}
\end{equation}
which is already diagonal with the complex eigenvalues 
\begin{align}
E_{1} &= \hbar \left[ \delta_{1} -\frac{i}{2} \left(R + \nu_1 \right) \right] \label{e:ev1} \\
E_{2} &= \hbar \left[\delta_{2} -\frac{i}{2} \left( \Gamma_{12} + \nu_{2} \right) \right]. \label{e:ev2}
\end{align}
Using these relations and equation (\ref{e:R}) for the incoherent pump rate $R$, we can express the shifts and decoherence rates via the eigenvalues of this effective Hamiltonian as:
\begin{equation}
\begin{split}
\delta_{1} &= \Re \{E_1\} / \hbar \\ 
\delta_{2} &= \Re \{E_2\} / \hbar \\
\nu_1 &= -2 \Im \{E_1\} / \hbar - R \\
\nu_2 &= -2 \Im \{E_2\} / \hbar - \Gamma_{12}
\end{split}
\label{e:expressions2level}
\end{equation}
Therefore, to reduce a driven multilevel system to an effective two-level system with incoherent pumping, we perform the following steps: First, we calculate the steady-state values for $\braket{\hat{\sigma}_{11}}$ and $\braket{\hat{\sigma}_{22}}$
, to obtain the effective repumping rate $R$ from equation~(\ref{e:R}). Second, we diagonalize the effective non-hermitian Hamiltonian (\ref{e:effHgeneral}) to get the complex eigenvalues $E_1$ and $E_2$. These eigenvalues correspond to the eigenstates with the highest overlap with the unperturbed clock states $\ket{1}$ and $\ket{2}$. Using these eigenvalues we calculate the shifts $\delta_1$ and $\delta_2$ and the decoherence rates $\nu_1$ and $\nu_2$ according to equation~(\ref{e:expressions2level}). 
In appendix~\ref{app:3-level} we apply this method analytically to a three-level system, and compare it to the "conventional" adiabatic elimination procedure. Note that the atoms are coupled to the cavity only on the weak $|1\rangle \leftrightarrow |2\rangle$ transition, which will not be adiabatically eliminated. Therefore the atom-cavity coupling of the reduced system is anyway retained and we can, in a good approximation, neglect the cavity field and perform the adiabatic elimination on a single atom. This also assumes that we neglected direct interaction between the atoms as well as any collective coupling of the atoms to the bath modes.

\section{Repumping scheme for bosonic strontium} 
\label{sec:repump_model}
\begin{figure}
\center
\includegraphics[width=0.9\columnwidth]{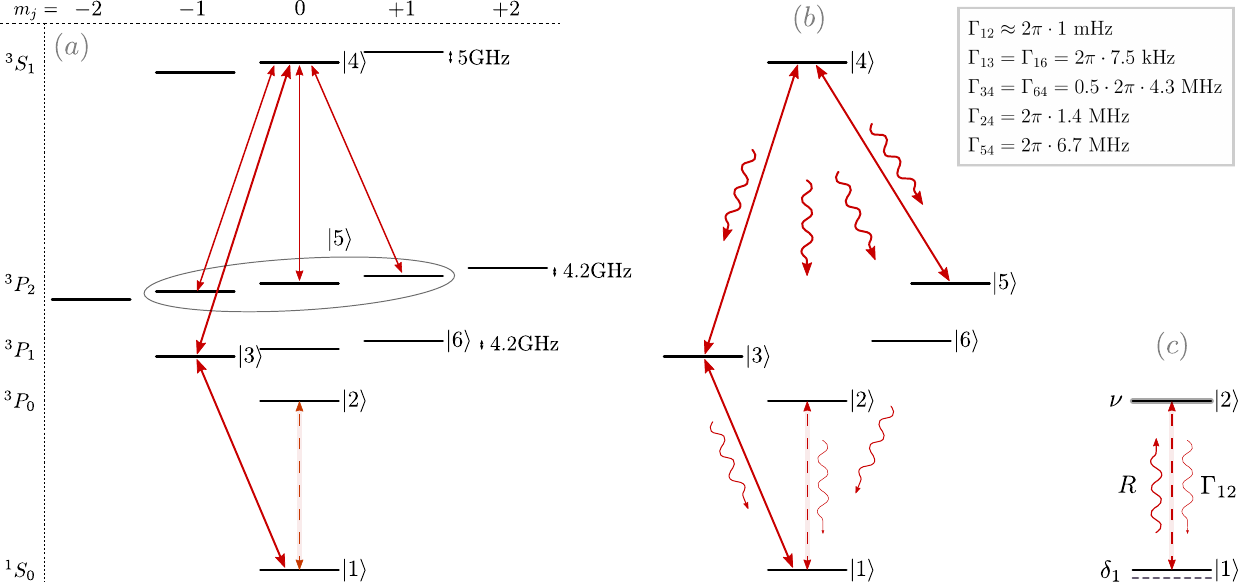}
\caption{\emph{Effective pump schemes:} Figure (a) shows all relevant atomic levels involved in CW pumping including the Zeeman sub-levels. Numbers show the level shifts for $B = 0.18\text{T}$. The directly involved transitions are indicated by the red solid lines. Figure (b) depicts a simplified six-level pump scheme with the relevant decay rates. (c) describes the resulting reduced two-level laser model including the effective emerging incoherent pump rate $R$ and ground state Stark shift $\delta_1$.  We also include an effective dephasing with rate $\nu$ on the lasing transition. The cavity coupling is indicated by the dashed line between $\ket{1}$ and $\ket{2}$. The decay rates are from \cite{NIST_ASD}.}
\label{fig:6level_pump}
\end{figure}

In the following we present the concrete proposed repumping scheme based on the actual level structure of ${ {}^{88}\text{Sr}}$, see figure \ref{fig:6level_pump}.  In order to allow for superradiant lasing we assume a fairly strong and homogeneous magnetic field $B = 0.18~\text{T}$ on the atoms, which induces an effective weak electric dipole coupling between the states $\ket{1}= {}^1\text{S}_0$ and $\ket{2}= {}^3\text{P}_0$ with spontaneous transition rate $\Gamma_{12} \approx 2 \pi \cdot 1~\text{mHz}$ \cite{taichenachev2006magnetic}. Naturally such a strong magnetic field splits the Zeeman sub-levels quite far, such that they can be addressed independently. To obtain sufficient population in the upper clock state $\ket{2} = {^3\text{P}_0}$ we consider the following processes: The atom is pumped coherently from the ground state $\ket{1} = {}^1\text{S}_0$ to $\ket{3} = {}^3\text{P}_{1,m=-1}$ and then further to $\ket{4} = {}^3{S}_{1,m=0}$. From there the atom can decay into the upper lasing state $\ket{2}$ as well as into the states $\ket{5} = {}^3\text{P}_{2, m=\{-1,0+1\}}$ and $\ket{6} = {}^3\text{P}_{1, m=-1}$. Note that the atoms can not decay from $\ket{4}$ to ${}^3\text{P}_{1, m=0}$, since this transition is forbidden by angular momentum selection rules. Furthermore we combine here the three relevant Zeeman sub-levels of the state ${}^3\text{P}_{2}$ to one state $\ket{5}$, this does not change the dynamics of the system, but one needs to be aware that in real experiments three individual lasers are needed to repump the atoms from these levels. 

The decay $\ket{4}\rightarrow \ket{2}$ is the desired final step in the excitation process. Since the state $\ket{5}$ has a lifetime even longer than the upper clock state, we need to additionally depopulate $\ket{5}$ to avoid trapping of too much population in this state. If the atoms decay into the state $\ket{6}$, we can either repump it back to the state $\ket{4}$, or simply let it decay further to the ground state. Pumping from the state $\ket{6}$ would further  increase the efficiency of the repumping process, but for simplicity we just consider spontaneous decay to the ground state. The Hamiltonian for the pumped six-level scheme (figure \ref{fig:6level_pump} (b)) in the rotating frame of the pump lasers then is
\begin{equation}
\hat{H}_\text{p} = - \Delta_{3}\sig_{33} - \Delta_{4}\sig_{44} - \Delta_{5}\sig_{55} + \Omega_{13}(\sig_{13} + \sig_{31}) + \Omega_{34}(\sig_{34} + \sig_{43}) + \Omega_{54} (\sig_{45} + \sig_{54}) 
\label{eq:H_pump}
\end{equation} 
with $\Delta_3 = \Delta_{13}$, $\Delta_4 = \Delta_3 + \Delta_{34}$ and $\Delta_5 = \Delta_4 - \Delta_{54}$. We define here $\Delta_{ij} = \omega_{ij}^l - \omega_{ij}$, $\omega_{ij}$ is the resonance frequency on the atomic transition $\ket{i} \leftrightarrow \ket{j}$, $\omega_{ij}^l$ is the frequency of the pump laser on this transition and $\Omega_{ij}$ the matrix element of the laser-induced transition. Dissipative processes for this pump scheme are described by the Liouvillian (\ref{e:LindbladGeneral}) with the parameters listed in table \ref{tab:jumps_pump}. These processes include all the relevant atomic decays, an effective phenomenological dephasing of the clock transition, as well as the dephasing induced by the pump lasers due to a finite linewidth. 
\begin{table}
\begin{center}
\begin{tabular}{ |c|c|c|c| } 
 \hline
 \# & jump & rate & description \\ 
 \hline
 1 & $\sig_{12}$ & $\Gamma_{12}$ & decay from $\ket{2}$ to $\ket{1}$ \\ 
 2 & $\sig_{13}$ & $\Gamma_{13}$ & decay from $\ket{3}$ to $\ket{1}$ \\
 3 & $\sig_{34}$ & $\Gamma_{34}$ & decay from $\ket{4}$ to $\ket{3}$ \\
 4 & $\sig_{24}$ & $\Gamma_{24}$ & decay from $\ket{4}$ to $\ket{2}$ \\
 5 & $\sig_{54}$ & $\Gamma_{54}$ & decay from $\ket{4}$ to $\ket{5}$ \\
 6 & $\sig_{64}$ & $\Gamma_{64}$ & decay from $\ket{4}$ to $\ket{6}$ \\
 7 & $\sig_{16}$ & $\Gamma_{16}$ & decay from $\ket{6}$ to $\ket{1}$ \\
 8 & $\sig_{22}$ & $\nu_{12}$ & general dephasing on $\ket{1} \leftrightarrow \ket{2}$ \\ 
 9 & $\sig_{33} + \sig_{44} + \sig_{55}$ & $\nu_{13}$ & pump laser linewidth on $\ket{1} \leftrightarrow \ket{3}$ \\
 10 & $\sig_{44} + \sig_{55}$ & $\nu_{34}$ & pump laser linewidth on $\ket{3} \leftrightarrow \ket{4}$ \\
 11 & $\sig_{55}$ & $\nu_{54}$ & pump laser linewidth on $\ket{4} \leftrightarrow \ket{5}$ \\
 \hline
\end{tabular}
\end{center}
\caption{\emph{Dissipative processes of the six-level pump scheme}.}
\label{tab:jumps_pump}
\end{table}

\subsection{Scanning over repumping parameters} 
\label{sec:opt_rep}

Using the method described in section \ref{sec:adiabatic} we analyze our Strontium six-level repumping scheme. For high atom numbers $N \gg 1$ an effective incoherent repumping rate $R$ above $\Gamma_{12}$ would already be sufficient for superradiant lasing \cite{meiser2009prospects, kazakov2016synchro}. A larger rate, however, leads to a higher output power and smaller linewidth, with an optimum at $R = 2 N g^2/\kappa$ \cite{meiser2009prospects}. Here $g$ is the atom cavity coupling constant and $\kappa$ the photon decay rate through the cavity mirrors. We will focus on incoherent repumping rates $R > 2 \pi \cdot 1~\mathrm{Hz}$, which is obviously much bigger than $\Gamma_{12} = 2 \pi \cdot 1~\mathrm{mHz}$ but not the optimum for usual atom numbers and cavity parameters. The issue with too high repumping rates is, that they usually require stronger pump fields which lead to bigger level shifts on the clock transition. Shifts per se would not be a problem if they are constant and known. However, due to uncertainties and fluctuations in the pump process, atoms at different positions might experience different shifts, which leads to an effective inhomogeneous broadening of the ensemble. But as long as the frequency distribution is small enough, the atoms can still synchronize and emit light collectively on a single narrow line \cite{xu2014synchronization, kazakov2016synchro, bychek2021superradiant}. For an ensemble with an inhomogeneous frequency broadening less than the incoherent pump rate $R$ the atoms synchronize in the superradiant regime.

Our aim is therefore to find parameters with an effective repumping rate $R > 2 \pi \cdot 1~\mathrm{Hz}$, but also sufficiently small frequency shift changes of the clock transition for realistic fluctuations and inaccuracies in the pump process. To this end we scan the effective repumping rate $R$ and the ground state shift $\delta_1$ on the relevant system parameters. Note that we only get shifts of the lower clock state in our model, since no pump laser couples to the upper clock state. 

A parameter set to achieve the above goal is: 
$\Omega_{13} = 2 \pi \cdot 1.5~{\rm kHz}$, 
$\Omega_{34} = 2 \pi \cdot 3.3~{\rm MHz}$, 
$\Omega_{54} = 2 \pi \cdot 100~{\rm kHz}$, 
$\Delta_{13} = -2 \pi \cdot 875~{\rm kHz}$,
$\Delta_{34} = -2 \pi \cdot 5~{\rm MHz} $ and 
$\Delta_{54} = -2 \pi \cdot 10~{\rm MHz}$. 
The corresponding two-level system parameters are $R \approx 2 \pi \cdot 1.91 ~\mathrm{Hz}$, $\delta_1 \approx 2 \pi \cdot 5.21~\mathrm{mHz}$ and $\nu \approx 2 \pi \cdot 3.93~\mathrm{Hz}$. These results are for pump laser linewidth of $\nu_{13} = \nu_{34} = \nu_{54} = 2 \pi \cdot 0.75~\mathrm{kHz}$ and a dephasing rate on the clock transition of $\nu = 2 \pi \cdot 1~\mathrm{Hz}$. We will use these parameters as our "standard" parameters, i.e. whenever parameters are kept constant in scans we use these. 

\begin{figure}
\center
\includegraphics[width=\columnwidth]{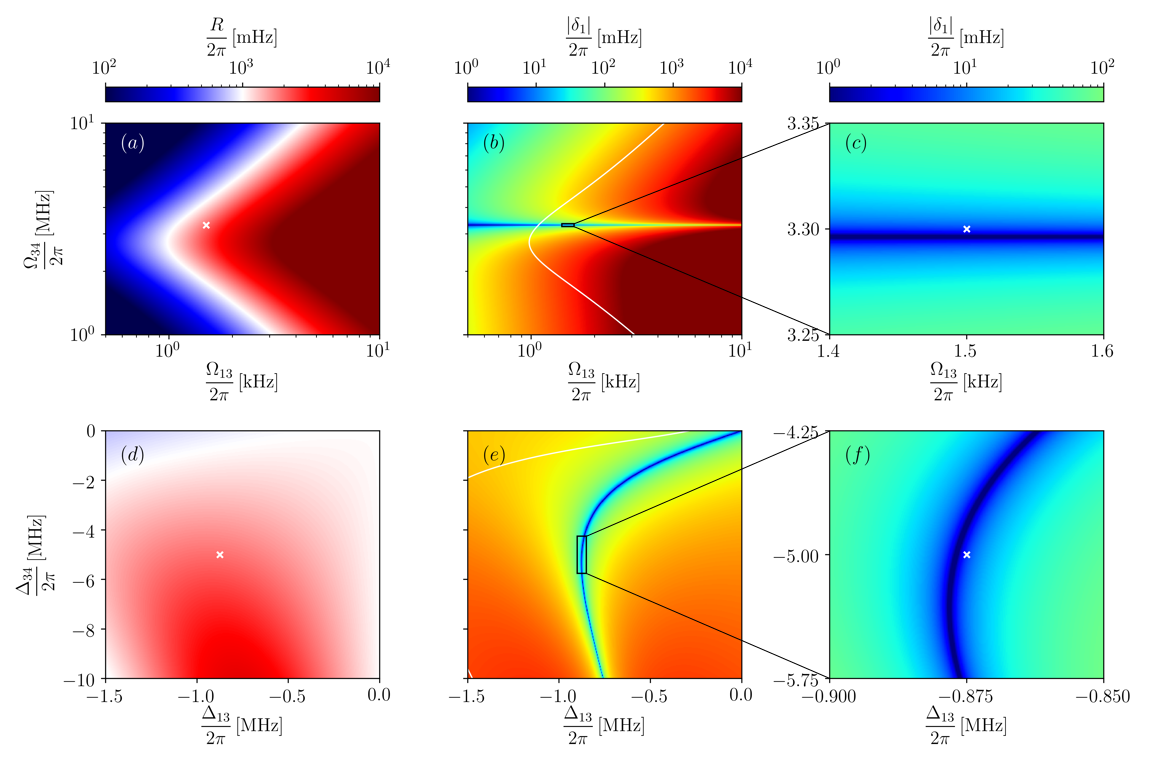}
\caption{\emph{Parameter scans  of the effective pump rate $R$ and the differential shift of the lasing transition $\delta_1$}. In the upper row (a-c) we vary the amplitudes of the two pump lasers and in the lower row (d-f) their frequency detuning. We are targeting regions of sufficiently high pump rate and very low shift. The white contour line in the $|\delta_1|$-scans depicts $R= 2\pi \cdot 1~\mathrm{Hz}$. Figure (c) and (f) show a zoom in a parameter region with small shifts $|\delta_1| < 2\pi \cdot 100~\mathrm{mHz}$. The white crosses indicate the parameters fixed in the other plots.}
\label{fig:2D-scans_3}
\end{figure}

Figure \ref{fig:2D-scans_3} shows the dependence of the effective repumping rate $R$ and the ground state shift $\delta_1$ on the Rabi-frequencies $\Omega_{13}$ and $\Omega_{34}$ (upper row) as well as on the detunings $\Delta_{13}$ and $\Delta_{34}$ (lower row), when the other parameters are kept constant. We do not show here scans on $\Delta_{54}$ and $\Omega_{54}$, since the dependences of $R$ and $\delta_1$ are very weak over a wide range of parameters, see one-dimensional scans in appendix \ref{app:1Dscans}. One can also see from these scans, that the dependence of both, $R$ and $\delta_1$, on $\Omega_{13}$ is quadratic. Therefore, the sensitivity of $\delta_1$ to variations of $\Omega_{13}$ is proportional to $\delta_1$, which means one should choose a working point with $\delta_1$ close to zero (dark blue regions in \fref{fig:2D-scans_3} (b) and (e)), otherwise rather small fluctuations on $\Omega_{13}$ might lead to big variations of $\delta_1$.

From the subplots (a) and (d) of \fref{fig:2D-scans_3} we can see, that there are wide regions with an effective repumping rate $R > 2 \pi \cdot 1~\mathrm{Hz}$. In the subplots (b) and (e) we plot the ground state shift and additionally indicate the relevant regions with the white line, which shows the repumping of $R > 2 \pi \cdot 1~\mathrm{Hz}$. In the panel (c) we zoom into an appropriate region for the Rabi-frequency scan. We see that for our parameters a deviation in $\Omega_{34}$ of $\pm 1.5 \%$ still has tolerable shifts. For $\Omega_{13}$ the suitable range is much bigger. 

In the panel (f) we show a proper region for the detuing scan. We chose an area with a suitable range for $\Delta_{13}$ of $2\pi \cdot 50~\mathrm{kHz}$ and for $\Delta_{34}$ of $2\pi \cdot 1.5~\mathrm{MHz}$. The reason to pick this region is the following: To avoid Doppler shifts a magic wavelength optical lattice is needed to trap the strontium atoms, but this lattice is in general only magic on the clock transition. This means for the other transitions the upper and lower state are not equally shifted and therefore non-clock transitions of atoms at different positions in the lattice have shifted resonance frequencies. However, the lattice can also be made magic on the $|1\rangle \leftrightarrow |3\rangle$ transition for a linearly polarized field, if one chooses the correct angle between the polarization axis and the quantization axis due to the static magnetic field \cite{Barker2016double}. Nonetheless, this does not work simultaneously on the $|3\rangle \leftrightarrow |4\rangle$ transition, which results in an effective inhomogeneous broadening of the transition frequency $\omega_{34}$ and hence in a $\Delta_{34}$ distribution. 
According to recent theoretical estimations \cite{Safronova21Private}, the scalar and tensor dynamic polarizabilities of the $(5s6s)^3S_1$ state at the 813~nm magic wavelength lattice are $\alpha_0(^3S_1) \approx -9\times 10^2~{\rm a.u.}$ (atomic units) and $\alpha_2(^3S_1) \approx 2~{\rm a.u.}$, respectively.
In turn, the scalar polarizabilities of the lasing states $\ket{1}$ and $\ket{2}$ at the magic wavelength are equal to $a_0(^1S_0)=a_0(^3P_0) \approx 2.8\times 10^2~{\rm a.u.}$ \cite{Martin13}. A thermal distribution of the atoms over different vibrational states and/or lattice sites with different potential depths will results into different shifts of the level $\ket{4}$. In particular, a temperature of $T=5~{\rm \mu K}$ corresponds to a shift of level $\ket{4}$ in the range of approximately 0.4~MHz, which directly results in a $\Delta_{34}$ distribution. Therefore we need to choose parameters with a wide suitable range for $\Delta_{34}$, but we can pick a point with a rather narrow range for $\Delta_{13}$.

In summary, the main result in this section is that it occurs to be possible to achieve a significant repumping rate $R$ together with a sufficient small and insensitive shift $\delta_1$. The optimal set of parameters will be individual for each experimental setup, but they can be found fast with the above described method. 

\section{Effective linewidth and shift in the reduced laser model}
\label{sec:laser}
In this section we calculate the spectrum of the superradiant laser, and demonstrate that our six-level laser model can be replaced by an effective two-level one. The laser model is given by $N$ identical six-level atoms pumped inside an optical cavity. In the rotating frame of the pump lasers and the unperturbed clock transition the Hamiltonian can be written as
\begin{align} \label{eq:H_l5}
H_\text{L6} =& -\Delta_\text{c} a^\dagger a + \sum_{k=1}^N \big[ - \Delta_{3}\sig^k_{33} - \Delta_{4}\sig^k_{44} - \Delta_{5}\sig^k_{55} + g (a^\dagger \sig^k_{12} + a \sig^k_{21}) \\ 
	+& \Omega_{13}(\sig^k_{13} + \sig^k_{31}) + \Omega_{34}(\sig^k_{34} + \sig^k_{43}) + \Omega_{54} (\sig^k_{45} + \sig^k_{54}) \big] . \notag
\end{align}
Here $\Delta_\text{c} = \omega_{12} - \omega_\text{c}$ is the detuning between the clock transition frequency and the cavity resonance frequency, and $g$ is the coupling coefficient between the cavity field and the clock transition. The dissipative processes for the atoms are listed in table \ref{tab:jumps_pump}. The dissipative processes of the cavity are according to table \ref{tab:jumps_cavity}. 
\begin{table}
\begin{center}
\begin{tabular}{ |c|c|c|c| } 
 \hline
 \# & jump & rate & description \\ 
 \hline
 1 & $a$ & $\kappa$ & cavity photon losses \\ 
 2 & $a^\dagger a$ & $\eta$ & fluctuations of the cavity resonance frequency \\
 \hline
\end{tabular}
\end{center}
\caption{\emph{Dissipative processes of the cavity field}.}
\label{tab:jumps_cavity}
\end{table}
Using that all atoms behave identically we derive second order cumulant equations \cite{Kubo1962generalized, plankensteiner2021quantumcumulantsjl} for the system variables and the correlation function \cite{6level_eqs}.
However, we can use the adiabatic elimination from section \ref{sec:adiabatic} to numerically reduce the six-level atom lasing model into an effective two-level atom lasing model. This simplifies the model drastically and increases the computational efficiency significantly. The Hamiltonian of this two-level lasing model is
\begin{equation} \label{eq:H_l2}
H_\text{L2} = -\Delta_\text{c} a^\dagger a + \sum_{k=1}^N \big[ - \delta^k_1 \sig^k_{22} + g_k (a^\dagger \sig^k_{12} + a \sig^k_{21}) \big]
\end{equation}
and the dissipative processes are given by table \ref{tab:jumps_cavity} for the cavity and table \ref{tab:2level} for each of the $N$ atoms individually. 

\begin{figure}
\center
\includegraphics[width=0.45\columnwidth]{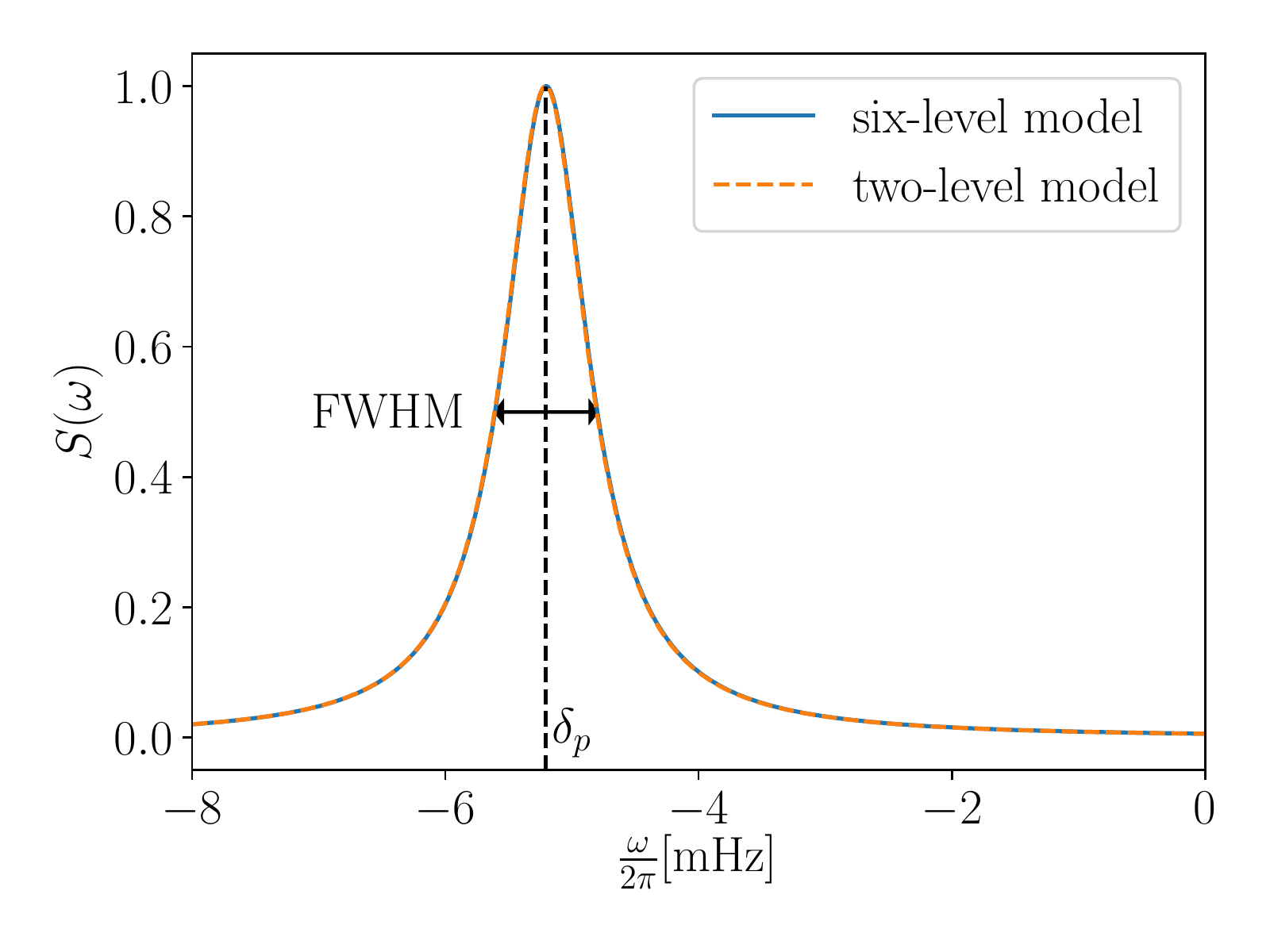}
\caption{\emph{Laser spectrum}. Comparison of the the effective two-level model with the full six-level model for a typical set of parameters. The laser properties are $\text{FWHM} = 2 \pi \cdot 0.806~\mathrm{mHz} \, (2 \pi \cdot 0.807~\mathrm{mHz} $ for the six-level model), $\delta_p = -2 \pi \cdot 5.20~\mathrm{mHz} \, (-2 \pi \cdot 5.21~\mathrm{mHz})$ and $n = 2.16 \, (2.15)$, with an inaccuracy of the effective model below $1 \%$. The atom number is $N = 2 \cdot 10^5$ and the cavity parameters are $\kappa = 2 \pi \cdot 75~\mathrm{kHz}$, $g = 2 \pi \cdot 2~\mathrm{Hz}$ and $\eta = 2 \pi \cdot 7.5~\mathrm{kHz}$.}
\label{fig:spectrum}
\end{figure}

In \fref{fig:spectrum} we see the excellent agreement of the laser properties calculated from the effective two-level lasing model and
the six-level model for our standard parameters. We compared these two models for many other relevant parameters, the laser properties (FWHM, $\delta_p, \, n$) always agreed well.

Finally, let us discuss the effect of the pump laser induced dephasing. The main mechanism of such a dephasing is that the atoms pumped into the upper states $\ket{3}$ and $\ket{4}$ can decay into the state $\ket{1}$, instead of into the upper lasing state $\ket{2}$, see also appendix~\ref{app:3-level}. Therefore, the dephasing rate $\nu$ needs to be proportional to the effective incoherent repumping rate $R$. Since the only way to get into the state $\ket{2}$ is to decay from $\ket{4}$ and since all other transitions are driven, the most prominent process to end up in $\ket{1}$ is via the decay into $\ket{6}$, therefore we can estimate
\begin{equation}
\nu	\approx R \frac{\Gamma_{64}}{\Gamma_{24}} + \nu_{12} \approx 1.5 R + \nu_{12}.
\end{equation}
For the parameters used in \fref{fig:spectrum} we get a FWHM of $\sim 2 \pi \cdot 0.81~\mathrm{mHz}$. In comparison the smallest linewidth we could theoretically get is $4g^2/\kappa \approx 2 \pi \cdot 0.21~\mathrm{mHz}$. With no dephasing at all ($\nu = 0$) this can, for these parameters, indeed be reached. Thus we see that the induced dephasing has an impact on the spectrum which is not to be neglected, but it is still reasonable. Note that the induced dephasing could e.g. be decreased in our case by additionally pumping the transition $|4\rangle \leftrightarrow |6\rangle$.

\section{Conclusions}

On the example of bosonic strontium trapped in a magic wavelength optical lattice we show that by choice of suitable pump laser parameters, it is possible to create significant population inversion on the clock transition with only a rather small shift and broadening of the lasing transition and the resulting active clock  line. In particular we found a parameter regime where the induced level shifts on the clock transition are small enough, such that the atoms can still synchronize and thus emit light collectively in the superradiant regime, where cavity noise plays no role. To perform our scans of the many parameters characterizing the complex multilevel system,  we have developed a fast numerical way to map the results to an effective two-level model, which can be well interpreted. For a range of generic test cases we have seen that the spectral and noise properties of these two models are in excellent agreement. The procedure can be adapted straight forward to find suitable repumping parameter for other alkaline-earth atoms. 

\section{Acknowledgment}

This work was supported by the European Union Horizon 2020 research and innovation programme, Quantum Flagship project No 820404 ``iqClock'' (C. H., D. P., G.K. and H. R.). Numerical simulations were performed with the open source frameworks QuantumOptics.jl~\cite{kramer2018quantumoptics} and QuantumCumulants.jl~\cite{plankensteiner2021quantumcumulantsjl}. The graphs were produced using the open source plotting library Matplotlib~\cite{hunter2007matplotlib}.
The authors thank  Marianna Safronova and Dmytro Filin for calculations of the dynamic polarizabilities on the $^3S_1$ state in neutral strontium atoms. 


\newpage

\appendix

\section{Analytic adiabatic elimination on a three-level atom} 
\label{app:3-level}
\begin{figure}[b]
\begin{center}
\resizebox{0.35\textwidth}{!}
{\includegraphics{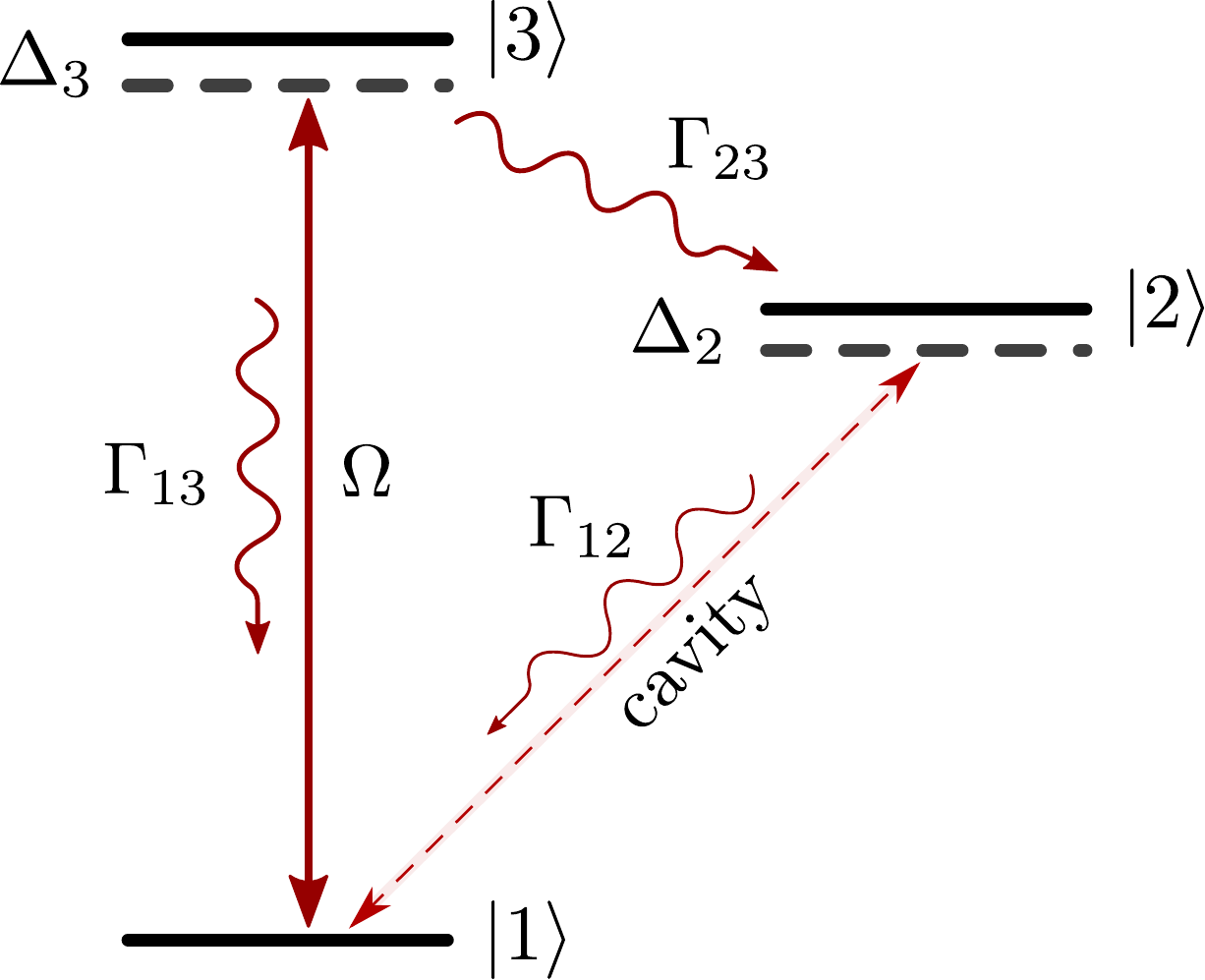}}
\end{center}
\caption{\emph{Three-level scheme.} The considered three-level scheme for the adiabatic elimination of level $\ket{3}$ is depicted. The atoms are coherently pumped on the transition $\ket{1} \leftrightarrow \ket{3}$ and the cavity couples on the transition $\ket{1} \leftrightarrow \ket{2}$.}
\label{fig:3level}
\end{figure}

Here we compare the "conventional" adaibatic elimination with the adiabatic elimination using the eigenvalues of the non-hermitian Hamiltonian. We apply the two methods analytically on a pumped three-level atom to eliminate the auxiliary level $\ket{3}$. 

The atom is coherently pumped on the transition $\ket{1} \leftrightarrow \ket{3}$ with a Rabi frequency $\Omega$ and laser detuning $\Delta_3$, see Fig.~\ref{fig:3level}. We suppose that the Rabi frequency $\Omega$ as well as the decay and decoherence rates associated with the state $\ket{3}$ ($\Gamma_{13}, \, \Gamma_{23}, \, \nu_3$) are much larger than all the other rates in the system, and that the total decay rate of level $\ket{3}$ is much larger than $\Omega$, therefore the population of the level $\ket{3}$ is much less than the populations of the levels $\ket{1}$ and $\ket{2}$. The Hamiltonian of a single atom can be written as
\begin{equation}
\H=\hbar (\Delta_2 \sig_{22} + \Delta_3 \sig_{33}) + \hbar \Omega (\sig_{13}+\sig_{31}),
\label{eq:2.B10}
\end{equation}
where $\Delta_2$ is some shift from the unperturbed atomic transition frequency. Jump operators and relaxation rates are listed in Table~\ref{tab:3level}. Note that we neglect the interaction with the weak cavity field, see section \ref{sec:adiabatic}.
\begin{table}
\begin{center}
\begin{tabular}{ |c|c|c|c| } 
 \hline
 \# & jump & rate & description \\ 
 \hline
 1 & $\sig_{12}$ & $\Gamma_{12}$ & decay from $\ket{2}$ to $\ket{1}$ \\ 
 2 & $\sig_{13}$ & $\Gamma_{13}$ & decay from $\ket{3}$ to $\ket{1}$ \\
 3 & $\sig_{23}$ & $\Gamma_{23}$ & decay from $\ket{3}$ to $\ket{2}$ \\
 4 & $\sig_{11}$ & $\nu_{1}^0$ & dephasing on $\ket{1}$ \\
 5 & $\sig_{22}$ & $\nu_{2}^0$ & dephasing on $\ket{2}$ \\
 6 & $\sig_{33}$ & $\nu_{3}^0$ & dephasing on $\ket{3}$ \\
\hline
\end{tabular}
\end{center}
\caption{{Dissipative processes of the three-level atom.}}
\label{tab:3level}
\end{table}
\subsection{"Conventional" adiabatic elimination}
First we perform an adiabatic elimination of the level $\ket{3}$, in the "conventional" way, similar to the one used in \cite{kazakov2017ions}. To this end we calculate the relevant equations for operator averages $\braket{\sig_{\alpha, \beta}}$, where $\alpha, \beta \in \{1,2,3\}$:
\begin{align}
\frac{d\braket{\sigee}}{dt}&= - \Gamma_{12} \braket{\sigee} + \Gamma_{23} \braket{\sigii}
\label{eq:2.B12}
\\
\frac{d\braket{\sigge}}{dt}&= -\left(\frac{\Gamma_{12}+ \nu_1^0+\nu_2^0}{2} + i \Delta_2 \right) \braket{\sigge}
 + i\Omega \braket{\sigie}
\label{eq:2.B13}
\\
\frac{d\braket{\sigii}}{dt}&= i\Omega \braket{ \siggi -\sigig } - \Gamma_3 \braket{\sigii}
\label{eq:2.B14}
\\
\frac{d\braket{\siggi}}{dt}&= -\left( 
 \frac{\Gamma_3+\nu_1^0+\nu_3^0}{2} + i\Delta_3 
 \right)\braket{\siggi}
 - i\Omega \braket{\siggg - \sigii}
\label{eq:2.B15}
\\
\frac{d\braket{\sigei}}{dt}&= -\left( 
 \frac{\Gamma_3+\Gamma_{12}+\nu_2^0+\nu_3^0}{2} + i\Delta_3 
 \right)\braket{\sigei}
 -i \Omega \braket{\sigeg}
\label{eq:2.B16} 
\end{align}
Here $\Gamma_3 = \Gamma_{13} + \Gamma_{23}$ is the total decay rate of the intermediate state $\ket{3}$. To perform the adiabatic elimination of $\braket{\sigii}$, $\braket{\sigei}$, $\braket{\siggi}$ we use $\Gamma_{12} \ll \Gamma_3$ and $\nu_1^0, \nu_2^0 \ll \nu_3^0$ as well as $\braket{\sigii} \ll \braket{\siggg}$. Introducing
\begin{equation}
\Gamma'=\frac{\Gamma_3+\nu_{3}^0}{2} \approx \frac{\Gamma_3+\Gamma_{12}+\nu_2^0+\nu_3^0}{2} \approx \frac{\Gamma_3+\nu_1^0+\nu_3^0}{2}
\label{eq:2.B17}
\end{equation}
we get 
\begin{align}
\braket{\siggi}&=\frac{-i \Omega}{\Gamma'+i \Delta_3} \braket{\siggg}
\label{eq:2.B18}
\\
\braket{\sigei}&=\frac{-i \Omega}{\Gamma'+i{\Delta_3\,}}\braket{\sigeg}
\label{eq:2.B19}
\\
\braket{\sigii}&=\frac{2 \Omega^2}{\Gamma_3} 
\frac{\Gamma'}{{\Gamma'}^2+{\Delta_3}^2}\braket{\siggg}.
\label{eq:2.B20}
\end{align}
Substituting these expressions into (\ref{eq:2.B12}) -- (\ref{eq:2.B13}), and introducing the repumping rate $R$, decoherence rate $\nu_{12}$ and effective shift $\Delta_{21}$ as
\begin{align}
R&= \frac{\Gamma_{23}}{\Gamma_3}
\frac{2 \Omega^2 \Gamma'}{{\Gamma'}^2+{\Delta_3}^2}
\label{eq:2.B21}
\\
\nu_{12}&=\nu_1^0+\nu_2^0+\frac{\Gamma_{13}}{\Gamma_3} \frac{2 \Omega^2 \Gamma'}{{\Gamma'}^2+{\Delta_3}^2}
\label{eq:2.B22}
\\
\Delta_{21}&= \Delta_2+\frac{\Omega^2 \Delta_3}{{\Gamma'}^2+{\Delta_3}^2}
\label{eq:2.B23}
\end{align}
we can rewrite the equations \eqref{eq:2.B12} and \eqref{eq:2.B13} as
\begin{align}
\frac{d\braket{\sigee}}{dt}&= - \Gamma_{12} \braket{\sigee} + R \braket{\siggg}
\label{eq:2.B24}
\\
\frac{d\braket{\sigge}}{dt}&= -\left(\frac{R+\Gamma_{12}+\nu_{12}}{2} + i \Delta_{21}
 \right) \braket{\sigge},
\label{eq:2.B25}
\end{align}
similar to eqs. \eqref{e:TwoLevelSig22} and \eqref{e:TwoLevelSig12}.

\subsection{Adiabatic elimination using the eigenvalues of the non-hermitian Hamiltonian}
Now we apply the procedure described in the end of Section \ref{sec:adiabatic}. The simplicity of the considered 3-level scheme allows to follow this method analytically. The expression for the repumping rate $R$, see eq. (\ref{e:R}), can be obtained from the steady-state expression of $\braket{\sig_{22}}$ and $\braket{\sig_{33}}$, 
\begin{align}
\braket{\sigee}&=\frac{\Gamma_{23}}{\Gamma_{12}}\braket{\sigii}=\frac{\Gamma_{23}}{\Gamma_{12}}\frac{2 \Omega^2 \Gamma'}{\Gamma(\Gamma'^2+{\Delta_3}^2)}\braket{\siggg}.
\label{eq:2.B26}
\end{align}
The result is the same as in (\ref{eq:2.B21}).

To determine the light shift and decoherence rate, one has to diagonalize the effective non-Hermitian Hamiltonian of our 3-level system in the absence of the cavity field. The Hamiltonian reads
\begin{equation}
\Heffnh = \H - \frac{i \hbar}{2} \sum_j R_j \Jdj \Jj
= \hbar (\delta_{2} \sig_{22} + \Delta_3 \sig_{33} + \Omega[\sig_{13}+\sig_{31}]) - \frac{i \hbar}{2} 
 \left[ 
 \Gamma_{12} \sig_{22} + \Gamma \sig_{33} 
 + \nu^0_{1} \sig_{11}+ \nu^0_{2} \sig_{22}+\nu^0_{3} \sig_{33}) 
 \right]
\label{eq:2.B27}
\end{equation}
with the eigenvalues
\begin{align}
\frac{E_2}{\hbar}&=\Delta_2-\frac{i}{2}(\Gamma_{12}+\nu_2^0)
\label{eq:2.B28} \\
\frac{E_{1,3}}{\hbar}&
 =\frac{\Delta_3-i\Gamma'}{2} 
 \left\{
 1 \mp \sqrt{ 1+
 \frac{4\Omega^2+2 i \nu_1^0 (\Delta_3-i\Gamma')}{(\Delta_3-i\Gamma')^2}} 
 \right\},
\label{eq:2.B29} 
\end{align}
where $\Gamma'$ is defined in (\ref{eq:2.B17}), and we neglected $\Gamma_{12}$, $\nu_1^0$ and $\nu_2^0$ in comparison with $\Gamma'$. For $\nu_1^0, \Omega \ll \Gamma'$ we can perform a Taylor expansion on the term $[4 \Omega^2+2 i \nu_1^0(\Delta_3-i\Gamma')]/(\Delta_3-i\Gamma')^2 \ll 1$ and find
\begin{align}
\frac{E_1}{\hbar}&\approx -\frac{\Omega^2 (\Delta_3+i \Gamma')}{{\Delta_3}^2+\Gamma'^2}-\frac{i \nu_1^0}{2}.
\label{eq:2.B30} 
\end{align}
Using expressions (\ref{e:expressions2level}), we get
\begin{align}
\Delta_1&=-\frac{\Omega^2 \Delta_3}{{\Gamma'}^2+{\Delta_3}^2}
\label{eq:2.B31} \\
\nu_{1}&= \nu_1^0
 + \frac{2 \Gamma' \Omega^2}{{\Gamma'}^2+{\Delta_3}^2}-R
 =\nu_1^0
 + \frac{\Gamma_{13}}{\Gamma}
 \frac{2 \Omega^2 \Gamma'}{{\Gamma'}^2+{\Delta_3}^2}.
\label{eq:2.B32}
\end{align}
Similarly, from (\ref{eq:2.B28}) follows $\nu_2=\nu_2^0$. Therefore we obtain
\begin{align}
\nu_{12}&=\nu_{1}+\nu_{2}=\nu_{1}^0+\nu_{2}^0+\frac{\Gamma_{13}}{\Gamma}\frac{2 \Omega^2 \Gamma'}{{\Gamma'}^2+{\Delta_3}^2}
\label{eq:2.B33}
\\
\Delta_{21}&=\Delta_2-\Delta_1=\Delta_2+\frac{\Omega^2 \Delta_3}{{\Gamma'}^2+{\Delta_3}^2},
\label{eq:2.B34}
\end{align}
this coincides with (\ref{eq:2.B22}) and (\ref{eq:2.B23}). Thus, we can see that th adiabatic elimination in such a 3-level system performed with the help of the diagonalization of the effective non-Hermitian Hamiltonian gives the same result as a "conventional" adiabatic elimination.

\section{One dimensional parameters scans} \label{app:1Dscans}
\begin{figure}
\center
\includegraphics[width=1.0\columnwidth]{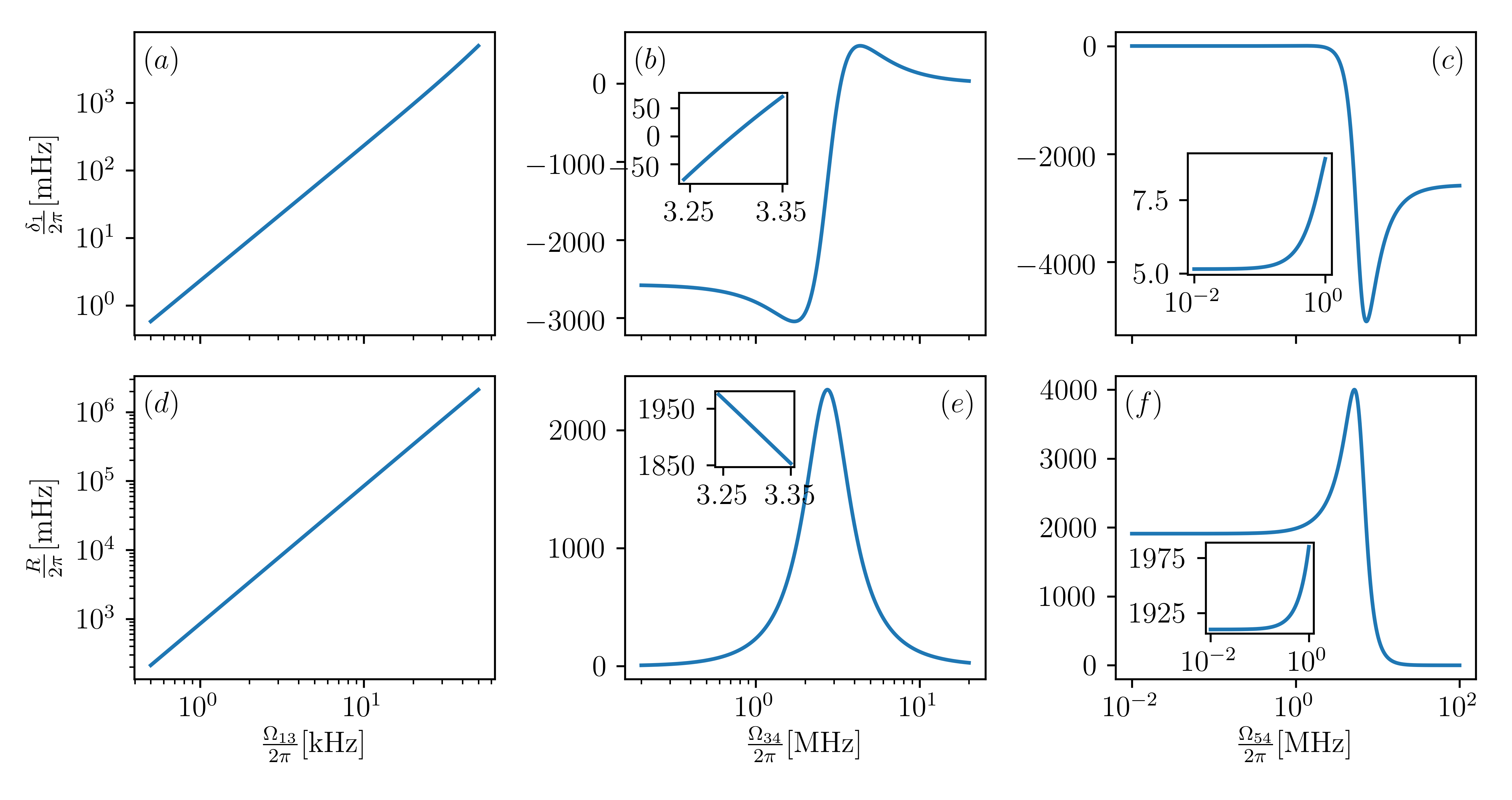}
\caption{\emph{Line shift (upper row) and pump rate (lower row) scans for varying pump amplitudes close to optimal operation conditions}. The dependence of the ground state shift and the repumping rate on the Rabi frequencies is shown with insets of interesting regions. The parameters when kept constant are our standard parameters from section \ref{sec:opt_rep}.}
\label{fig:mini-scans_Omega}
\end{figure}

\begin{figure}
\center
\includegraphics[width=1.0\columnwidth]{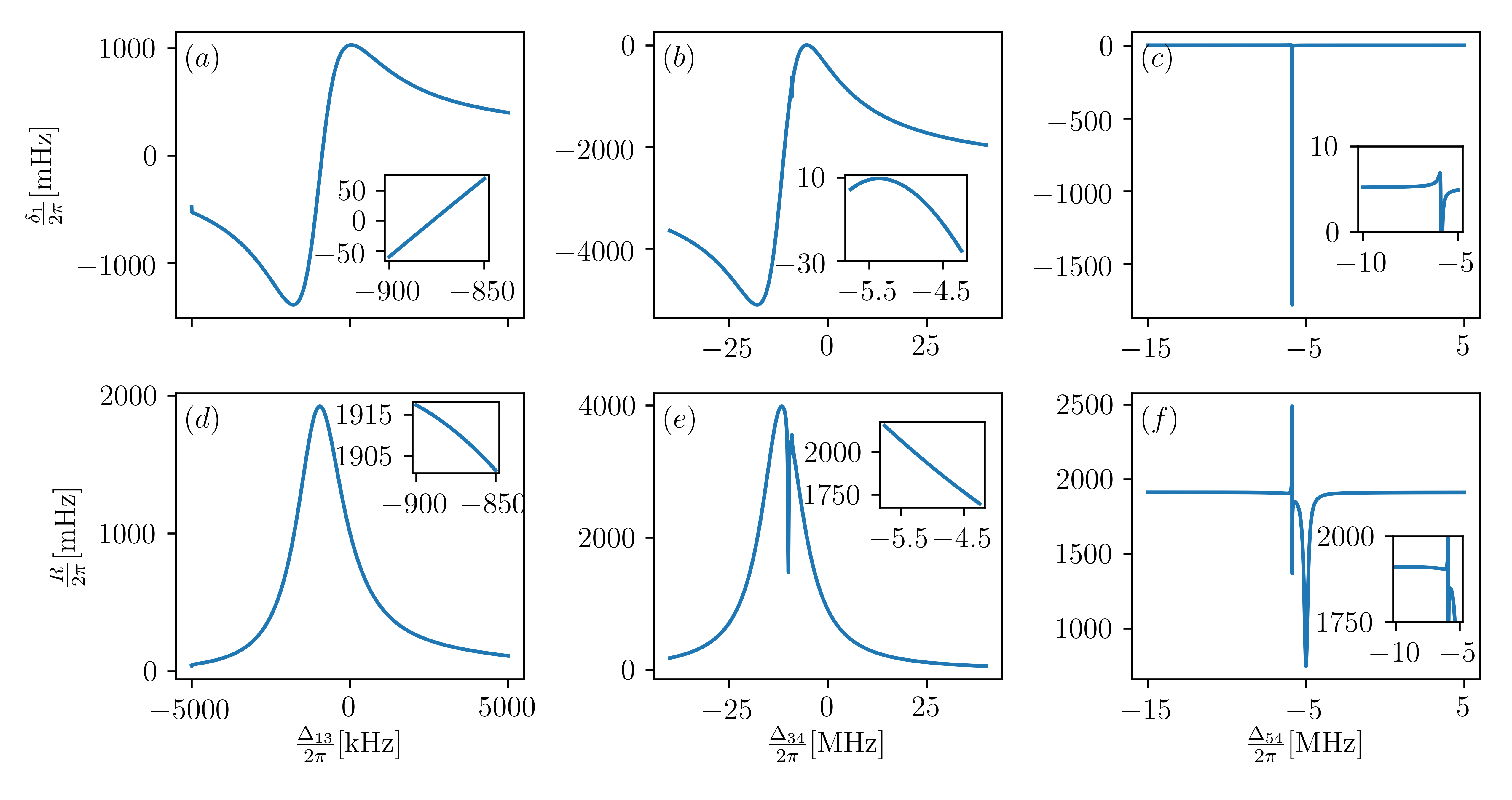}
\caption{\emph{Line shift (upper row) and pump rate (lower row) scans for varying pump detunings close to optimal operation conditions}. The dependence of the ground state shift and the repumping rate on the detuings is shown with insets of interesting regions. The parameters when kept constant are our standard parameters from section \ref{sec:opt_rep}.}
\label{fig:mini-scans_Delta}
\end{figure}

To get a better insight of the dependence on the different repumping parameters we show here one dimensional scans. \fref{fig:mini-scans_Omega} shows scans on the Rabi-frequencies and \fref{fig:mini-scans_Delta} on the detunings for $R$ and $\delta_1$. 

The scans on $\Omega_{13}$, \fref{fig:mini-scans_Omega} (a) and (d), show a quadratic dependence of $R$ and $\delta_1$ on the relevant regions of $\Omega_{13}$ with a constant prefactor depending on the other system parameters. A proper choice of parameters can reduce the prefactor of the ground state shift by orders of magnitudes, while the pump rate prefactor stays almost the same. For the $\Omega_{34}$-scans, \fref{fig:mini-scans_Omega} (b) and (e), we find only a relative small area at approximately $\Omega_{34} \approx 2\pi \cdot 3.3~\mathrm{MHz}$ (see inset) where a repumping rate $R > 2\pi \cdot 1~\mathrm{Hz}$ can be achieved and the range of the shift $| \delta_1 |$ is sufficiently small. For $\Omega_{54}$, \fref{fig:mini-scans_Omega} (c) and (f), on the other hand, all values below $2\pi \cdot 1~\mathrm{MHz}$ have an almost constant $R$ and $\delta_1$. But note that for smaller values of $\Omega_{54}$ more population is trapped in $\ket{5}$, this is undesired since less population will contribute to lasing. However, for our parameters this gets only relevant for $\Omega_{54}  < 2\pi \cdot 1~\mathrm{kHz}$.

As we know from section \ref{sec:opt_rep} $\Delta_{13}$ can be controlled much more precisely than $\Delta_{34}$ due to the magic wavelength lattice, therefore we choose a parameter regime in which changes of $\Delta_{34}$ are far less important. \fref{fig:mini-scans_Delta} illustrates this very well. Differences in $\Delta_{13}$ of $\pm 2\pi \cdot 25~\mathrm{kHz}$ lead to shifts of $\pm 2\pi \cdot 50~\mathrm{mHz}$, whereas changes of $\pm 2\pi \cdot 0.75~\mathrm{MHz}$ in $\Delta_{34}$ lead only to shifts in a range of approximately $2\pi \cdot 40~\mathrm{mHz}$, see \fref{fig:mini-scans_Delta} (a) and (b) respectively. 
For the $\Delta_{54}$ dependency, \fref{fig:mini-scans_Delta} (c) and (f), we find that $R$ and $\delta_1$ do not significantly change for detunings between $-2\pi \cdot 15$ and $-2\pi \cdot 6~\mathrm{MHz}$. Thus $\Delta_{54}$ does not need to be precisely controlled.

\bibliography{5-level-laserRef}

\begin{thebibliography}{52}%
\makeatletter
\providecommand \@ifxundefined [1]{%
 \@ifx{#1\undefined}
}%
\providecommand \@ifnum [1]{%
 \ifnum #1\expandafter \@firstoftwo
 \else \expandafter \@secondoftwo
 \fi
}%
\providecommand \@ifx [1]{%
 \ifx #1\expandafter \@firstoftwo
 \else \expandafter \@secondoftwo
 \fi
}%
\providecommand \natexlab [1]{#1}%
\providecommand \enquote  [1]{``#1''}%
\providecommand \bibnamefont  [1]{#1}%
\providecommand \bibfnamefont [1]{#1}%
\providecommand \citenamefont [1]{#1}%
\providecommand \href@noop [0]{\@secondoftwo}%
\providecommand \href [0]{\begingroup \@sanitize@url \@href}%
\providecommand \@href[1]{\@@startlink{#1}\@@href}%
\providecommand \@@href[1]{\endgroup#1\@@endlink}%
\providecommand \@sanitize@url [0]{\catcode `\\12\catcode `\$12\catcode
  `\&12\catcode `\#12\catcode `\^12\catcode `\_12\catcode `\%12\relax}%
\providecommand \@@startlink[1]{}%
\providecommand \@@endlink[0]{}%
\providecommand \url  [0]{\begingroup\@sanitize@url \@url }%
\providecommand \@url [1]{\endgroup\@href {#1}{\urlprefix }}%
\providecommand \urlprefix  [0]{URL }%
\providecommand \Eprint [0]{\href }%
\providecommand \doibase [0]{http://dx.doi.org/}%
\providecommand \selectlanguage [0]{\@gobble}%
\providecommand \bibinfo  [0]{\@secondoftwo}%
\providecommand \bibfield  [0]{\@secondoftwo}%
\providecommand \translation [1]{[#1]}%
\providecommand \BibitemOpen [0]{}%
\providecommand \bibitemStop [0]{}%
\providecommand \bibitemNoStop [0]{.\EOS\space}%
\providecommand \EOS [0]{\spacefactor3000\relax}%
\providecommand \BibitemShut  [1]{\csname bibitem#1\endcsname}%
\let\auto@bib@innerbib\@empty
\bibitem [{\citenamefont {Oelker}\ \emph {et~al.}(2019)\citenamefont {Oelker},
  \citenamefont {Hutson}, \citenamefont {Kennedy}, \citenamefont {Sonderhouse},
  \citenamefont {Bothwell}, \citenamefont {Goban}, \citenamefont {Kedar},
  \citenamefont {Sanner}, \citenamefont {Robinson}, \citenamefont {Marti},
  \citenamefont {Matei}, \citenamefont {Legero}, \citenamefont {Giunta},
  \citenamefont {Holzwarth}, \citenamefont {Riehle}, \citenamefont {Sterr},\
  and\ \citenamefont {Ye}}]{Oelker2019}%
  \BibitemOpen
  \bibfield  {author} {\bibinfo {author} {\bibfnamefont {E.}~\bibnamefont
  {Oelker}}, \bibinfo {author} {\bibfnamefont {R.~B.}\ \bibnamefont {Hutson}},
  \bibinfo {author} {\bibfnamefont {C.~J.}\ \bibnamefont {Kennedy}}, \bibinfo
  {author} {\bibfnamefont {L.}~\bibnamefont {Sonderhouse}}, \bibinfo {author}
  {\bibfnamefont {T.}~\bibnamefont {Bothwell}}, \bibinfo {author}
  {\bibfnamefont {A.}~\bibnamefont {Goban}}, \bibinfo {author} {\bibfnamefont
  {D.}~\bibnamefont {Kedar}}, \bibinfo {author} {\bibfnamefont
  {C.}~\bibnamefont {Sanner}}, \bibinfo {author} {\bibfnamefont {J.~M.}\
  \bibnamefont {Robinson}}, \bibinfo {author} {\bibfnamefont {G.~E.}\
  \bibnamefont {Marti}}, \bibinfo {author} {\bibfnamefont {D.~G.}\ \bibnamefont
  {Matei}}, \bibinfo {author} {\bibfnamefont {T.}~\bibnamefont {Legero}},
  \bibinfo {author} {\bibfnamefont {M.}~\bibnamefont {Giunta}}, \bibinfo
  {author} {\bibfnamefont {R.}~\bibnamefont {Holzwarth}}, \bibinfo {author}
  {\bibfnamefont {F.}~\bibnamefont {Riehle}}, \bibinfo {author} {\bibfnamefont
  {U.}~\bibnamefont {Sterr}}, \ and\ \bibinfo {author} {\bibfnamefont
  {J.}~\bibnamefont {Ye}},\ }\href {\doibase 10.1038/s41566-019-0493-4}
  {\bibfield  {journal} {\bibinfo  {journal} {Nature Photonics}\ }\textbf
  {\bibinfo {volume} {13}},\ \bibinfo {pages} {714} (\bibinfo {year}
  {2019})}\BibitemShut {NoStop}%
\bibitem [{\citenamefont {Numata}\ \emph {et~al.}(2004)\citenamefont {Numata},
  \citenamefont {Kemery},\ and\ \citenamefont {Camp}}]{numata2004thermal}%
  \BibitemOpen
  \bibfield  {author} {\bibinfo {author} {\bibfnamefont {K.}~\bibnamefont
  {Numata}}, \bibinfo {author} {\bibfnamefont {A.}~\bibnamefont {Kemery}}, \
  and\ \bibinfo {author} {\bibfnamefont {J.}~\bibnamefont {Camp}},\ }\href
  {https://link.aps.org/doi/10.1103/PhysRevLett.93.250602} {\bibfield
  {journal} {\bibinfo  {journal} {Physical review letters}\ }\textbf {\bibinfo
  {volume} {93}},\ \bibinfo {pages} {250602} (\bibinfo {year}
  {2004})}\BibitemShut {NoStop}%
\bibitem [{\citenamefont {Chen}(2009)}]{Chen2009active}%
  \BibitemOpen
  \bibfield  {author} {\bibinfo {author} {\bibfnamefont {J.}~\bibnamefont
  {Chen}},\ }\href {https://doi.org/10.1007/s11434-009-0073-y} {\bibfield
  {journal} {\bibinfo  {journal} {Chinese Science Bulletin}\ }\textbf {\bibinfo
  {volume} {54}},\ \bibinfo {pages} {348} (\bibinfo {year} {2009})}\BibitemShut
  {NoStop}%
\bibitem [{\citenamefont {Meiser}\ \emph {et~al.}(2009)\citenamefont {Meiser},
  \citenamefont {Ye}, \citenamefont {Carlson},\ and\ \citenamefont
  {Holland}}]{meiser2009prospects}%
  \BibitemOpen
  \bibfield  {author} {\bibinfo {author} {\bibfnamefont {D.}~\bibnamefont
  {Meiser}}, \bibinfo {author} {\bibfnamefont {J.}~\bibnamefont {Ye}}, \bibinfo
  {author} {\bibfnamefont {D.}~\bibnamefont {Carlson}}, \ and\ \bibinfo
  {author} {\bibfnamefont {M.}~\bibnamefont {Holland}},\ }\href
  {https://link.aps.org/doi/10.1103/PhysRevLett.102.163601} {\bibfield
  {journal} {\bibinfo  {journal} {Physical review letters}\ }\textbf {\bibinfo
  {volume} {102}},\ \bibinfo {pages} {163601} (\bibinfo {year}
  {2009})}\BibitemShut {NoStop}%
\bibitem [{\citenamefont {Bohnet}\ \emph {et~al.}(2012)\citenamefont {Bohnet},
  \citenamefont {Chen}, \citenamefont {Weiner}, \citenamefont {Meiser},
  \citenamefont {Holland},\ and\ \citenamefont {Thompson}}]{bohnet2012steady}%
  \BibitemOpen
  \bibfield  {author} {\bibinfo {author} {\bibfnamefont {J.~G.}\ \bibnamefont
  {Bohnet}}, \bibinfo {author} {\bibfnamefont {Z.}~\bibnamefont {Chen}},
  \bibinfo {author} {\bibfnamefont {J.~M.}\ \bibnamefont {Weiner}}, \bibinfo
  {author} {\bibfnamefont {D.}~\bibnamefont {Meiser}}, \bibinfo {author}
  {\bibfnamefont {M.~J.}\ \bibnamefont {Holland}}, \ and\ \bibinfo {author}
  {\bibfnamefont {J.~K.}\ \bibnamefont {Thompson}},\ }\href
  {https://doi.org/10.1038/nature10920} {\bibfield  {journal} {\bibinfo
  {journal} {Nature}\ }\textbf {\bibinfo {volume} {484}},\ \bibinfo {pages}
  {78} (\bibinfo {year} {2012})}\BibitemShut {NoStop}%
\bibitem [{\citenamefont {Bohnet}\ \emph {et~al.}(2014)\citenamefont {Bohnet},
  \citenamefont {Chen}, \citenamefont {Weiner}, \citenamefont {Cox},\ and\
  \citenamefont {Thompson}}]{bohnet2014linear}%
  \BibitemOpen
  \bibfield  {author} {\bibinfo {author} {\bibfnamefont {J.~G.}\ \bibnamefont
  {Bohnet}}, \bibinfo {author} {\bibfnamefont {Z.}~\bibnamefont {Chen}},
  \bibinfo {author} {\bibfnamefont {J.~M.}\ \bibnamefont {Weiner}}, \bibinfo
  {author} {\bibfnamefont {K.~C.}\ \bibnamefont {Cox}}, \ and\ \bibinfo
  {author} {\bibfnamefont {J.~K.}\ \bibnamefont {Thompson}},\ }\href {\doibase
  10.1103/PhysRevA.89.013806} {\bibfield  {journal} {\bibinfo  {journal} {Phys.
  Rev. A}\ }\textbf {\bibinfo {volume} {89}},\ \bibinfo {pages} {013806}
  (\bibinfo {year} {2014})}\BibitemShut {NoStop}%
\bibitem [{\citenamefont {Bychek}\ \emph {et~al.}(2021)\citenamefont {Bychek},
  \citenamefont {Hotter}, \citenamefont {Plankensteiner},\ and\ \citenamefont
  {Ritsch}}]{bychek2021superradiant}%
  \BibitemOpen
  \bibfield  {author} {\bibinfo {author} {\bibfnamefont {A.}~\bibnamefont
  {Bychek}}, \bibinfo {author} {\bibfnamefont {C.}~\bibnamefont {Hotter}},
  \bibinfo {author} {\bibfnamefont {D.}~\bibnamefont {Plankensteiner}}, \ and\
  \bibinfo {author} {\bibfnamefont {H.}~\bibnamefont {Ritsch}},\ }\href
  {https://doi.org/10.12688/openreseurope.13781.1} {\bibfield  {journal}
  {\bibinfo  {journal} {Open Research Europe 2021, 1:73}\ } (\bibinfo {year}
  {2021})}\BibitemShut {NoStop}%
\bibitem [{\citenamefont {Debnath}\ \emph {et~al.}(2018)\citenamefont
  {Debnath}, \citenamefont {Zhang},\ and\ \citenamefont
  {M{\o}lmer}}]{debnath2018lasing}%
  \BibitemOpen
  \bibfield  {author} {\bibinfo {author} {\bibfnamefont {K.}~\bibnamefont
  {Debnath}}, \bibinfo {author} {\bibfnamefont {Y.}~\bibnamefont {Zhang}}, \
  and\ \bibinfo {author} {\bibfnamefont {K.}~\bibnamefont {M{\o}lmer}},\ }\href
  {https://link.aps.org/doi/10.1103/PhysRevA.98.063837} {\bibfield  {journal}
  {\bibinfo  {journal} {Physical Review A}\ }\textbf {\bibinfo {volume} {98}},\
  \bibinfo {pages} {063837} (\bibinfo {year} {2018})}\BibitemShut {NoStop}%
\bibitem [{\citenamefont {Gogyan}\ \emph {et~al.}(2020)\citenamefont {Gogyan},
  \citenamefont {Kazakov}, \citenamefont {Bober},\ and\ \citenamefont
  {Zawada}}]{Gogyan2020characterisation}%
  \BibitemOpen
  \bibfield  {author} {\bibinfo {author} {\bibfnamefont {A.}~\bibnamefont
  {Gogyan}}, \bibinfo {author} {\bibfnamefont {G.}~\bibnamefont {Kazakov}},
  \bibinfo {author} {\bibfnamefont {M.}~\bibnamefont {Bober}}, \ and\ \bibinfo
  {author} {\bibfnamefont {M.}~\bibnamefont {Zawada}},\ }\href {\doibase
  10.1364/OE.381991} {\bibfield  {journal} {\bibinfo  {journal} {Opt. Express}\
  }\textbf {\bibinfo {volume} {28}},\ \bibinfo {pages} {6881} (\bibinfo {year}
  {2020})}\BibitemShut {NoStop}%
\bibitem [{\citenamefont {Haake}\ \emph {et~al.}(1993)\citenamefont {Haake},
  \citenamefont {Kolobov}, \citenamefont {Fabre}, \citenamefont {Giacobino},\
  and\ \citenamefont {Reynaud}}]{Haake1993superradiant}%
  \BibitemOpen
  \bibfield  {author} {\bibinfo {author} {\bibfnamefont {F.}~\bibnamefont
  {Haake}}, \bibinfo {author} {\bibfnamefont {M.~I.}\ \bibnamefont {Kolobov}},
  \bibinfo {author} {\bibfnamefont {C.}~\bibnamefont {Fabre}}, \bibinfo
  {author} {\bibfnamefont {E.}~\bibnamefont {Giacobino}}, \ and\ \bibinfo
  {author} {\bibfnamefont {S.}~\bibnamefont {Reynaud}},\ }\href {\doibase
  10.1103/PhysRevLett.71.995} {\bibfield  {journal} {\bibinfo  {journal} {Phys.
  Rev. Lett.}\ }\textbf {\bibinfo {volume} {71}},\ \bibinfo {pages} {995}
  (\bibinfo {year} {1993})}\BibitemShut {NoStop}%
\bibitem [{\citenamefont {Hotter}\ \emph {et~al.}(2019)\citenamefont {Hotter},
  \citenamefont {Plankensteiner}, \citenamefont {Ostermann},\ and\
  \citenamefont {Ritsch}}]{hotter2019superradiant}%
  \BibitemOpen
  \bibfield  {author} {\bibinfo {author} {\bibfnamefont {C.}~\bibnamefont
  {Hotter}}, \bibinfo {author} {\bibfnamefont {D.}~\bibnamefont
  {Plankensteiner}}, \bibinfo {author} {\bibfnamefont {L.}~\bibnamefont
  {Ostermann}}, \ and\ \bibinfo {author} {\bibfnamefont {H.}~\bibnamefont
  {Ritsch}},\ }\href
  {http://www.osapublishing.org/oe/abstract.cfm?URI=oe-27-22-31193} {\bibfield
  {journal} {\bibinfo  {journal} {Optics express}\ }\textbf {\bibinfo {volume}
  {27}},\ \bibinfo {pages} {31193} (\bibinfo {year} {2019})}\BibitemShut
  {NoStop}%
\bibitem [{\citenamefont {J\"ager}\ \emph {et~al.}(2021)\citenamefont
  {J\"ager}, \citenamefont {Liu}, \citenamefont {Shankar}, \citenamefont
  {Cooper},\ and\ \citenamefont {Holland}}]{Jaeger21}%
  \BibitemOpen
  \bibfield  {author} {\bibinfo {author} {\bibfnamefont {S.~B.}\ \bibnamefont
  {J\"ager}}, \bibinfo {author} {\bibfnamefont {H.}~\bibnamefont {Liu}},
  \bibinfo {author} {\bibfnamefont {A.}~\bibnamefont {Shankar}}, \bibinfo
  {author} {\bibfnamefont {J.}~\bibnamefont {Cooper}}, \ and\ \bibinfo {author}
  {\bibfnamefont {M.~J.}\ \bibnamefont {Holland}},\ }\href {\doibase
  10.1103/PhysRevA.103.013720} {\bibfield  {journal} {\bibinfo  {journal}
  {Phys. Rev. A}\ }\textbf {\bibinfo {volume} {103}},\ \bibinfo {pages}
  {013720} (\bibinfo {year} {2021})}\BibitemShut {NoStop}%
\bibitem [{\citenamefont {Kazakov}\ and\ \citenamefont
  {Schumm}(2013)}]{kazakov2013active}%
  \BibitemOpen
  \bibfield  {author} {\bibinfo {author} {\bibfnamefont {G.~A.}\ \bibnamefont
  {Kazakov}}\ and\ \bibinfo {author} {\bibfnamefont {T.}~\bibnamefont
  {Schumm}},\ }\href {https://link.aps.org/doi/10.1103/PhysRevA.87.013821}
  {\bibfield  {journal} {\bibinfo  {journal} {Physical Review A}\ }\textbf
  {\bibinfo {volume} {87}},\ \bibinfo {pages} {013821} (\bibinfo {year}
  {2013})}\BibitemShut {NoStop}%
\bibitem [{\citenamefont {Kazakov}\ and\ \citenamefont
  {Schumm}(2014)}]{kazakov2014active}%
  \BibitemOpen
  \bibfield  {author} {\bibinfo {author} {\bibfnamefont {G.~A.}\ \bibnamefont
  {Kazakov}}\ and\ \bibinfo {author} {\bibfnamefont {T.}~\bibnamefont
  {Schumm}},\ }in\ \href@noop {} {\emph {\bibinfo {booktitle} {2014 European
  Frequency and Time Forum (EFTF)}}}\ (\bibinfo {organization} {IEEE},\
  \bibinfo {year} {2014})\ pp.\ \bibinfo {pages} {411--414}\BibitemShut
  {NoStop}%
\bibitem [{\citenamefont {Kazakov}\ and\ \citenamefont
  {Schumm}(2017)}]{kazakov2016synchro}%
  \BibitemOpen
  \bibfield  {author} {\bibinfo {author} {\bibfnamefont {G.~A.}\ \bibnamefont
  {Kazakov}}\ and\ \bibinfo {author} {\bibfnamefont {T.}~\bibnamefont
  {Schumm}},\ }\href {\doibase 10.1103/PhysRevA.95.023839} {\bibfield
  {journal} {\bibinfo  {journal} {Phys. Rev. A}\ }\textbf {\bibinfo {volume}
  {95}},\ \bibinfo {pages} {023839} (\bibinfo {year} {2017})}\BibitemShut
  {NoStop}%
\bibitem [{\citenamefont {Kazakov}\ \emph {et~al.}(2017)\citenamefont
  {Kazakov}, \citenamefont {Bohnet},\ and\ \citenamefont
  {Schumm}}]{kazakov2017ions}%
  \BibitemOpen
  \bibfield  {author} {\bibinfo {author} {\bibfnamefont {G.~A.}\ \bibnamefont
  {Kazakov}}, \bibinfo {author} {\bibfnamefont {J.}~\bibnamefont {Bohnet}}, \
  and\ \bibinfo {author} {\bibfnamefont {T.}~\bibnamefont {Schumm}},\ }\href
  {\doibase 10.1103/PhysRevA.96.023412} {\bibfield  {journal} {\bibinfo
  {journal} {Phys. Rev. A}\ }\textbf {\bibinfo {volume} {96}},\ \bibinfo
  {pages} {023412} (\bibinfo {year} {2017})}\BibitemShut {NoStop}%
\bibitem [{\citenamefont {Laske}\ \emph {et~al.}(2019)\citenamefont {Laske},
  \citenamefont {Winter},\ and\ \citenamefont {Hemmerich}}]{laske2019pulse}%
  \BibitemOpen
  \bibfield  {author} {\bibinfo {author} {\bibfnamefont {T.}~\bibnamefont
  {Laske}}, \bibinfo {author} {\bibfnamefont {H.}~\bibnamefont {Winter}}, \
  and\ \bibinfo {author} {\bibfnamefont {A.}~\bibnamefont {Hemmerich}},\ }\href
  {https://link.aps.org/doi/10.1103/PhysRevLett.123.103601} {\bibfield
  {journal} {\bibinfo  {journal} {Physical Review Letters}\ }\textbf {\bibinfo
  {volume} {123}},\ \bibinfo {pages} {103601} (\bibinfo {year}
  {2019})}\BibitemShut {NoStop}%
\bibitem [{\citenamefont {Liu}\ \emph {et~al.}(2020)\citenamefont {Liu},
  \citenamefont {J{\"a}ger}, \citenamefont {Yu}, \citenamefont {Touzard},
  \citenamefont {Shankar}, \citenamefont {Holland},\ and\ \citenamefont
  {Nicholson}}]{liu2020rugged}%
  \BibitemOpen
  \bibfield  {author} {\bibinfo {author} {\bibfnamefont {H.}~\bibnamefont
  {Liu}}, \bibinfo {author} {\bibfnamefont {S.~B.}\ \bibnamefont {J{\"a}ger}},
  \bibinfo {author} {\bibfnamefont {X.}~\bibnamefont {Yu}}, \bibinfo {author}
  {\bibfnamefont {S.}~\bibnamefont {Touzard}}, \bibinfo {author} {\bibfnamefont
  {A.}~\bibnamefont {Shankar}}, \bibinfo {author} {\bibfnamefont {M.~J.}\
  \bibnamefont {Holland}}, \ and\ \bibinfo {author} {\bibfnamefont {T.~L.}\
  \bibnamefont {Nicholson}},\ }\href
  {https://link.aps.org/doi/10.1103/PhysRevLett.125.253602} {\bibfield
  {journal} {\bibinfo  {journal} {Physical Review Letters}\ }\textbf {\bibinfo
  {volume} {125}},\ \bibinfo {pages} {253602} (\bibinfo {year}
  {2020})}\BibitemShut {NoStop}%
\bibitem [{\citenamefont {Maier}\ \emph {et~al.}(2014)\citenamefont {Maier},
  \citenamefont {Kraemer}, \citenamefont {Ostermann},\ and\ \citenamefont
  {Ritsch}}]{maier2014superradiant}%
  \BibitemOpen
  \bibfield  {author} {\bibinfo {author} {\bibfnamefont {T.}~\bibnamefont
  {Maier}}, \bibinfo {author} {\bibfnamefont {S.}~\bibnamefont {Kraemer}},
  \bibinfo {author} {\bibfnamefont {L.}~\bibnamefont {Ostermann}}, \ and\
  \bibinfo {author} {\bibfnamefont {H.}~\bibnamefont {Ritsch}},\ }\href
  {http://www.osapublishing.org/oe/abstract.cfm?URI=oe-22-11-13269} {\bibfield
  {journal} {\bibinfo  {journal} {Optics express}\ }\textbf {\bibinfo {volume}
  {22}},\ \bibinfo {pages} {13269} (\bibinfo {year} {2014})}\BibitemShut
  {NoStop}%
\bibitem [{\citenamefont {Meiser}\ and\ \citenamefont
  {Holland}(2010{\natexlab{a}})}]{meiser2010intensity}%
  \BibitemOpen
  \bibfield  {author} {\bibinfo {author} {\bibfnamefont {D.}~\bibnamefont
  {Meiser}}\ and\ \bibinfo {author} {\bibfnamefont {M.}~\bibnamefont
  {Holland}},\ }\href {https://link.aps.org/doi/10.1103/PhysRevA.81.063827}
  {\bibfield  {journal} {\bibinfo  {journal} {Physical Review A}\ }\textbf
  {\bibinfo {volume} {81}},\ \bibinfo {pages} {063827} (\bibinfo {year}
  {2010}{\natexlab{a}})}\BibitemShut {NoStop}%
\bibitem [{\citenamefont {Meiser}\ and\ \citenamefont
  {Holland}(2010{\natexlab{b}})}]{meiser2010steady}%
  \BibitemOpen
  \bibfield  {author} {\bibinfo {author} {\bibfnamefont {D.}~\bibnamefont
  {Meiser}}\ and\ \bibinfo {author} {\bibfnamefont {M.}~\bibnamefont
  {Holland}},\ }\href {https://link.aps.org/doi/10.1103/PhysRevA.81.033847}
  {\bibfield  {journal} {\bibinfo  {journal} {Physical Review A}\ }\textbf
  {\bibinfo {volume} {81}},\ \bibinfo {pages} {033847} (\bibinfo {year}
  {2010}{\natexlab{b}})}\BibitemShut {NoStop}%
\bibitem [{\citenamefont {Norcia}\ and\ \citenamefont
  {Thompson}(2016)}]{norcia2016cold}%
  \BibitemOpen
  \bibfield  {author} {\bibinfo {author} {\bibfnamefont {M.~A.}\ \bibnamefont
  {Norcia}}\ and\ \bibinfo {author} {\bibfnamefont {J.~K.}\ \bibnamefont
  {Thompson}},\ }\href {https://link.aps.org/doi/10.1103/PhysRevX.6.011025}
  {\bibfield  {journal} {\bibinfo  {journal} {Physical Review X}\ }\textbf
  {\bibinfo {volume} {6}},\ \bibinfo {pages} {011025} (\bibinfo {year}
  {2016})}\BibitemShut {NoStop}%
\bibitem [{\citenamefont {Norcia}\ \emph {et~al.}(2016)\citenamefont {Norcia},
  \citenamefont {Winchester}, \citenamefont {Cline},\ and\ \citenamefont
  {Thompson}}]{norcia2016superradiance}%
  \BibitemOpen
  \bibfield  {author} {\bibinfo {author} {\bibfnamefont {M.~A.}\ \bibnamefont
  {Norcia}}, \bibinfo {author} {\bibfnamefont {M.~N.}\ \bibnamefont
  {Winchester}}, \bibinfo {author} {\bibfnamefont {J.~R.}\ \bibnamefont
  {Cline}}, \ and\ \bibinfo {author} {\bibfnamefont {J.~K.}\ \bibnamefont
  {Thompson}},\ }\href {https://www.science.org/doi/abs/10.1126/sciadv.1601231}
  {\bibfield  {journal} {\bibinfo  {journal} {Science advances}\ }\textbf
  {\bibinfo {volume} {2}},\ \bibinfo {pages} {e1601231} (\bibinfo {year}
  {2016})}\BibitemShut {NoStop}%
\bibitem [{\citenamefont {Norcia}\ \emph
  {et~al.}(2018{\natexlab{a}})\citenamefont {Norcia}, \citenamefont
  {Lewis-Swan}, \citenamefont {Cline}, \citenamefont {Zhu}, \citenamefont
  {Rey},\ and\ \citenamefont {Thompson}}]{norcia2018cavity}%
  \BibitemOpen
  \bibfield  {author} {\bibinfo {author} {\bibfnamefont {M.~A.}\ \bibnamefont
  {Norcia}}, \bibinfo {author} {\bibfnamefont {R.~J.}\ \bibnamefont
  {Lewis-Swan}}, \bibinfo {author} {\bibfnamefont {J.~R.}\ \bibnamefont
  {Cline}}, \bibinfo {author} {\bibfnamefont {B.}~\bibnamefont {Zhu}}, \bibinfo
  {author} {\bibfnamefont {A.~M.}\ \bibnamefont {Rey}}, \ and\ \bibinfo
  {author} {\bibfnamefont {J.~K.}\ \bibnamefont {Thompson}},\ }\href
  {https://www.science.org/doi/abs/10.1126/science.aar3102} {\bibfield
  {journal} {\bibinfo  {journal} {Science}\ }\textbf {\bibinfo {volume}
  {361}},\ \bibinfo {pages} {259} (\bibinfo {year}
  {2018}{\natexlab{a}})}\BibitemShut {NoStop}%
\bibitem [{\citenamefont {Norcia}\ \emph
  {et~al.}(2018{\natexlab{b}})\citenamefont {Norcia}, \citenamefont {Cline},
  \citenamefont {Muniz}, \citenamefont {Robinson}, \citenamefont {Hutson},
  \citenamefont {Goban}, \citenamefont {Marti}, \citenamefont {Ye},\ and\
  \citenamefont {Thompson}}]{norcia2018frequency}%
  \BibitemOpen
  \bibfield  {author} {\bibinfo {author} {\bibfnamefont {M.~A.}\ \bibnamefont
  {Norcia}}, \bibinfo {author} {\bibfnamefont {J.~R.}\ \bibnamefont {Cline}},
  \bibinfo {author} {\bibfnamefont {J.~A.}\ \bibnamefont {Muniz}}, \bibinfo
  {author} {\bibfnamefont {J.~M.}\ \bibnamefont {Robinson}}, \bibinfo {author}
  {\bibfnamefont {R.~B.}\ \bibnamefont {Hutson}}, \bibinfo {author}
  {\bibfnamefont {A.}~\bibnamefont {Goban}}, \bibinfo {author} {\bibfnamefont
  {G.~E.}\ \bibnamefont {Marti}}, \bibinfo {author} {\bibfnamefont
  {J.}~\bibnamefont {Ye}}, \ and\ \bibinfo {author} {\bibfnamefont {J.~K.}\
  \bibnamefont {Thompson}},\ }\href
  {https://link.aps.org/doi/10.1103/PhysRevX.8.021036} {\bibfield  {journal}
  {\bibinfo  {journal} {Physical Review X}\ }\textbf {\bibinfo {volume} {8}},\
  \bibinfo {pages} {021036} (\bibinfo {year} {2018}{\natexlab{b}})}\BibitemShut
  {NoStop}%
\bibitem [{\citenamefont {Sch{\"a}ffer}\ \emph {et~al.}(2020)\citenamefont
  {Sch{\"a}ffer}, \citenamefont {Tang}, \citenamefont {Henriksen},
  \citenamefont {J{\o}rgensen}, \citenamefont {Christensen},\ and\
  \citenamefont {Thomsen}}]{schaffer2020lasing}%
  \BibitemOpen
  \bibfield  {author} {\bibinfo {author} {\bibfnamefont {S.~A.}\ \bibnamefont
  {Sch{\"a}ffer}}, \bibinfo {author} {\bibfnamefont {M.}~\bibnamefont {Tang}},
  \bibinfo {author} {\bibfnamefont {M.~R.}\ \bibnamefont {Henriksen}}, \bibinfo
  {author} {\bibfnamefont {A.~A.}\ \bibnamefont {J{\o}rgensen}}, \bibinfo
  {author} {\bibfnamefont {B.~T.}\ \bibnamefont {Christensen}}, \ and\ \bibinfo
  {author} {\bibfnamefont {J.~W.}\ \bibnamefont {Thomsen}},\ }\href
  {https://link.aps.org/doi/10.1103/PhysRevA.101.013819} {\bibfield  {journal}
  {\bibinfo  {journal} {Physical Review a}\ }\textbf {\bibinfo {volume}
  {101}},\ \bibinfo {pages} {013819} (\bibinfo {year} {2020})}\BibitemShut
  {NoStop}%
\bibitem [{\citenamefont {Tang}\ \emph {et~al.}(2021)\citenamefont {Tang},
  \citenamefont {Sch{\"a}ffer}, \citenamefont {J{\o}rgensen}, \citenamefont
  {Henriksen}, \citenamefont {Christensen}, \citenamefont {M{\"u}ller},\ and\
  \citenamefont {Thomsen}}]{tang2021cavity}%
  \BibitemOpen
  \bibfield  {author} {\bibinfo {author} {\bibfnamefont {M.}~\bibnamefont
  {Tang}}, \bibinfo {author} {\bibfnamefont {S.~A.}\ \bibnamefont
  {Sch{\"a}ffer}}, \bibinfo {author} {\bibfnamefont {A.~A.}\ \bibnamefont
  {J{\o}rgensen}}, \bibinfo {author} {\bibfnamefont {M.~R.}\ \bibnamefont
  {Henriksen}}, \bibinfo {author} {\bibfnamefont {B.~T.}\ \bibnamefont
  {Christensen}}, \bibinfo {author} {\bibfnamefont {J.~H.}\ \bibnamefont
  {M{\"u}ller}}, \ and\ \bibinfo {author} {\bibfnamefont {J.~W.}\ \bibnamefont
  {Thomsen}},\ }\href
  {https://link.aps.org/doi/10.1103/PhysRevResearch.3.033258} {\bibfield
  {journal} {\bibinfo  {journal} {Physical Review Research}\ }\textbf {\bibinfo
  {volume} {3}},\ \bibinfo {pages} {033258} (\bibinfo {year}
  {2021})}\BibitemShut {NoStop}%
\bibitem [{\citenamefont {Shankar}\ \emph {et~al.}(2021)\citenamefont
  {Shankar}, \citenamefont {Reilly}, \citenamefont {J{\"a}ger},\ and\
  \citenamefont {Holland}}]{shankar2021subradiant}%
  \BibitemOpen
  \bibfield  {author} {\bibinfo {author} {\bibfnamefont {A.}~\bibnamefont
  {Shankar}}, \bibinfo {author} {\bibfnamefont {J.~T.}\ \bibnamefont {Reilly}},
  \bibinfo {author} {\bibfnamefont {S.~B.}\ \bibnamefont {J{\"a}ger}}, \ and\
  \bibinfo {author} {\bibfnamefont {M.~J.}\ \bibnamefont {Holland}},\
  }\href@noop {} {\bibfield  {journal} {\bibinfo  {journal} {arXiv preprint
  arXiv:2103.07402}\ } (\bibinfo {year} {2021})}\BibitemShut {NoStop}%
\bibitem [{\citenamefont {Weiner}\ \emph {et~al.}(2017)\citenamefont {Weiner},
  \citenamefont {Cox}, \citenamefont {Bohnet},\ and\ \citenamefont
  {Thompson}}]{weiner2017phase}%
  \BibitemOpen
  \bibfield  {author} {\bibinfo {author} {\bibfnamefont {J.~M.}\ \bibnamefont
  {Weiner}}, \bibinfo {author} {\bibfnamefont {K.~C.}\ \bibnamefont {Cox}},
  \bibinfo {author} {\bibfnamefont {J.~G.}\ \bibnamefont {Bohnet}}, \ and\
  \bibinfo {author} {\bibfnamefont {J.~K.}\ \bibnamefont {Thompson}},\ }\href
  {https://link.aps.org/doi/10.1103/PhysRevA.95.033808} {\bibfield  {journal}
  {\bibinfo  {journal} {Physical Review A}\ }\textbf {\bibinfo {volume} {95}},\
  \bibinfo {pages} {033808} (\bibinfo {year} {2017})}\BibitemShut {NoStop}%
\bibitem [{\citenamefont {Xu}\ \emph {et~al.}(2014)\citenamefont {Xu},
  \citenamefont {Tieri}, \citenamefont {Fine}, \citenamefont {Thompson},\ and\
  \citenamefont {Holland}}]{xu2014synchronization}%
  \BibitemOpen
  \bibfield  {author} {\bibinfo {author} {\bibfnamefont {M.}~\bibnamefont
  {Xu}}, \bibinfo {author} {\bibfnamefont {D.~A.}\ \bibnamefont {Tieri}},
  \bibinfo {author} {\bibfnamefont {E.}~\bibnamefont {Fine}}, \bibinfo {author}
  {\bibfnamefont {J.~K.}\ \bibnamefont {Thompson}}, \ and\ \bibinfo {author}
  {\bibfnamefont {M.~J.}\ \bibnamefont {Holland}},\ }\href
  {https://link.aps.org/doi/10.1103/PhysRevLett.113.154101} {\bibfield
  {journal} {\bibinfo  {journal} {Physical review letters}\ }\textbf {\bibinfo
  {volume} {113}},\ \bibinfo {pages} {154101} (\bibinfo {year}
  {2014})}\BibitemShut {NoStop}%
\bibitem [{\citenamefont {Zhang}\ \emph {et~al.}(2021)\citenamefont {Zhang},
  \citenamefont {Shan},\ and\ \citenamefont
  {M{\o}lmer}}]{zhang2021ultranarrow}%
  \BibitemOpen
  \bibfield  {author} {\bibinfo {author} {\bibfnamefont {Y.}~\bibnamefont
  {Zhang}}, \bibinfo {author} {\bibfnamefont {C.}~\bibnamefont {Shan}}, \ and\
  \bibinfo {author} {\bibfnamefont {K.}~\bibnamefont {M{\o}lmer}},\ }\href
  {https://link.aps.org/doi/10.1103/PhysRevLett.126.123602} {\bibfield
  {journal} {\bibinfo  {journal} {Physical Review Letters}\ }\textbf {\bibinfo
  {volume} {126}},\ \bibinfo {pages} {123602} (\bibinfo {year}
  {2021})}\BibitemShut {NoStop}%
\bibitem [{\citenamefont {Goldenberg}\ \emph {et~al.}(1960)\citenamefont
  {Goldenberg}, \citenamefont {Kleppner},\ and\ \citenamefont
  {Ramsey}}]{goldenberg1960atomic}%
  \BibitemOpen
  \bibfield  {author} {\bibinfo {author} {\bibfnamefont {H.~M.}\ \bibnamefont
  {Goldenberg}}, \bibinfo {author} {\bibfnamefont {D.}~\bibnamefont
  {Kleppner}}, \ and\ \bibinfo {author} {\bibfnamefont {N.~F.}\ \bibnamefont
  {Ramsey}},\ }\href {\doibase 10.1103/PhysRevLett.5.361} {\bibfield  {journal}
  {\bibinfo  {journal} {Phys. Rev. Lett.}\ }\textbf {\bibinfo {volume} {5}},\
  \bibinfo {pages} {361} (\bibinfo {year} {1960})}\BibitemShut {NoStop}%
\bibitem [{\citenamefont {Strelnitski}\ \emph {et~al.}(1995)\citenamefont
  {Strelnitski}, \citenamefont {Ponomarev},\ and\ \citenamefont
  {Smith}}]{strelnitski1995hydrogen}%
  \BibitemOpen
  \bibfield  {author} {\bibinfo {author} {\bibfnamefont {V.~S.}\ \bibnamefont
  {Strelnitski}}, \bibinfo {author} {\bibfnamefont {V.~O.}\ \bibnamefont
  {Ponomarev}}, \ and\ \bibinfo {author} {\bibfnamefont {H.~A.}\ \bibnamefont
  {Smith}},\ }\href@noop {} {\bibfield  {journal} {\bibinfo  {journal} {arXiv
  preprint astro-ph/9511118}\ } (\bibinfo {year} {1995})}\BibitemShut {NoStop}%
\bibitem [{\citenamefont {Salzburger}\ and\ \citenamefont
  {Ritsch}(2007)}]{salzburger2007atom}%
  \BibitemOpen
  \bibfield  {author} {\bibinfo {author} {\bibfnamefont {T.}~\bibnamefont
  {Salzburger}}\ and\ \bibinfo {author} {\bibfnamefont {H.}~\bibnamefont
  {Ritsch}},\ }\href {\doibase 10.1103/PhysRevA.75.061601} {\bibfield
  {journal} {\bibinfo  {journal} {Phys. Rev. A}\ }\textbf {\bibinfo {volume}
  {75}},\ \bibinfo {pages} {061601} (\bibinfo {year} {2007})}\BibitemShut
  {NoStop}%
\bibitem [{\citenamefont {Chen}\ \emph {et~al.}(2019)\citenamefont {Chen},
  \citenamefont {Bennetts}, \citenamefont {Escudero}, \citenamefont
  {Pasquiou},\ and\ \citenamefont {Schreck}}]{chen2019continuous}%
  \BibitemOpen
  \bibfield  {author} {\bibinfo {author} {\bibfnamefont {C.-C.}\ \bibnamefont
  {Chen}}, \bibinfo {author} {\bibfnamefont {S.}~\bibnamefont {Bennetts}},
  \bibinfo {author} {\bibfnamefont {R.~G.}\ \bibnamefont {Escudero}}, \bibinfo
  {author} {\bibfnamefont {B.}~\bibnamefont {Pasquiou}}, \ and\ \bibinfo
  {author} {\bibfnamefont {F.}~\bibnamefont {Schreck}},\ }\href {\doibase
  10.1103/PhysRevApplied.12.044014} {\bibfield  {journal} {\bibinfo  {journal}
  {Phys. Rev. Applied}\ }\textbf {\bibinfo {volume} {12}},\ \bibinfo {pages}
  {044014} (\bibinfo {year} {2019})}\BibitemShut {NoStop}%
\bibitem [{\citenamefont {Katori}\ \emph {et~al.}(2003)\citenamefont {Katori},
  \citenamefont {Takamoto}, \citenamefont {Pal’Chikov},\ and\ \citenamefont
  {Ovsiannikov}}]{katori2003ultrastable}%
  \BibitemOpen
  \bibfield  {author} {\bibinfo {author} {\bibfnamefont {H.}~\bibnamefont
  {Katori}}, \bibinfo {author} {\bibfnamefont {M.}~\bibnamefont {Takamoto}},
  \bibinfo {author} {\bibfnamefont {V.}~\bibnamefont {Pal’Chikov}}, \ and\
  \bibinfo {author} {\bibfnamefont {V.}~\bibnamefont {Ovsiannikov}},\ }\href
  {https://link.aps.org/doi/10.1103/PhysRevLett.91.173005} {\bibfield
  {journal} {\bibinfo  {journal} {Physical Review Letters}\ }\textbf {\bibinfo
  {volume} {91}},\ \bibinfo {pages} {173005} (\bibinfo {year}
  {2003})}\BibitemShut {NoStop}%
\bibitem [{\citenamefont {Ye}\ \emph {et~al.}(2008)\citenamefont {Ye},
  \citenamefont {Kimble},\ and\ \citenamefont {Katori}}]{ye2008quantum}%
  \BibitemOpen
  \bibfield  {author} {\bibinfo {author} {\bibfnamefont {J.}~\bibnamefont
  {Ye}}, \bibinfo {author} {\bibfnamefont {H.}~\bibnamefont {Kimble}}, \ and\
  \bibinfo {author} {\bibfnamefont {H.}~\bibnamefont {Katori}},\ }\href
  {https://www.science.org/doi/abs/10.1126/science.1148259} {\bibfield
  {journal} {\bibinfo  {journal} {science}\ }\textbf {\bibinfo {volume}
  {320}},\ \bibinfo {pages} {1734} (\bibinfo {year} {2008})}\BibitemShut
  {NoStop}%
\bibitem [{\citenamefont {Escudero}\ \emph {et~al.}(2021)\citenamefont
  {Escudero}, \citenamefont {Chen}, \citenamefont {Bennetts}, \citenamefont
  {Pasquiou},\ and\ \citenamefont {Schreck}}]{escudero2021steady}%
  \BibitemOpen
  \bibfield  {author} {\bibinfo {author} {\bibfnamefont {R.~G.}\ \bibnamefont
  {Escudero}}, \bibinfo {author} {\bibfnamefont {C.-C.}\ \bibnamefont {Chen}},
  \bibinfo {author} {\bibfnamefont {S.}~\bibnamefont {Bennetts}}, \bibinfo
  {author} {\bibfnamefont {B.}~\bibnamefont {Pasquiou}}, \ and\ \bibinfo
  {author} {\bibfnamefont {F.}~\bibnamefont {Schreck}},\ }\href@noop {}
  {\bibfield  {journal} {\bibinfo  {journal} {arXiv preprint arXiv:2104.06814}\
  } (\bibinfo {year} {2021})}\BibitemShut {NoStop}%
\bibitem [{\citenamefont {Bennetts}\ \emph {et~al.}(2017)\citenamefont
  {Bennetts}, \citenamefont {Chen}, \citenamefont {Pasquiou}, \citenamefont
  {Schreck} \emph {et~al.}}]{bennetts2017steady}%
  \BibitemOpen
  \bibfield  {author} {\bibinfo {author} {\bibfnamefont {S.}~\bibnamefont
  {Bennetts}}, \bibinfo {author} {\bibfnamefont {C.-C.}\ \bibnamefont {Chen}},
  \bibinfo {author} {\bibfnamefont {B.}~\bibnamefont {Pasquiou}}, \bibinfo
  {author} {\bibfnamefont {F.}~\bibnamefont {Schreck}},  \emph {et~al.},\
  }\href {https://link.aps.org/doi/10.1103/PhysRevLett.119.223202} {\bibfield
  {journal} {\bibinfo  {journal} {Physical review letters}\ }\textbf {\bibinfo
  {volume} {119}},\ \bibinfo {pages} {223202} (\bibinfo {year}
  {2017})}\BibitemShut {NoStop}%
\bibitem [{\citenamefont {Kubo}(1962)}]{Kubo1962generalized}%
  \BibitemOpen
  \bibfield  {author} {\bibinfo {author} {\bibfnamefont {R.}~\bibnamefont
  {Kubo}},\ }\href {https://doi.org/10.1143/JPSJ.17.1100} {\bibfield  {journal}
  {\bibinfo  {journal} {Journal of the Physical Society of Japan}\ }\textbf
  {\bibinfo {volume} {17}},\ \bibinfo {pages} {1100} (\bibinfo {year}
  {1962})}\BibitemShut {NoStop}%
\bibitem [{\citenamefont {Plankensteiner}\ \emph {et~al.}(2021)\citenamefont
  {Plankensteiner}, \citenamefont {Hotter},\ and\ \citenamefont
  {Ritsch}}]{plankensteiner2021quantumcumulantsjl}%
  \BibitemOpen
  \bibfield  {author} {\bibinfo {author} {\bibfnamefont {D.}~\bibnamefont
  {Plankensteiner}}, \bibinfo {author} {\bibfnamefont {C.}~\bibnamefont
  {Hotter}}, \ and\ \bibinfo {author} {\bibfnamefont {H.}~\bibnamefont
  {Ritsch}},\ }\href@noop {} {\enquote {\bibinfo {title} {Quantumcumulants.jl:
  A julia framework for generalized mean-field equations in open quantum
  systems},}\ } (\bibinfo {year} {2021}),\ \Eprint
  {http://arxiv.org/abs/2105.01657} {arXiv:2105.01657 [quant-ph]} \BibitemShut
  {NoStop}%
\bibitem [{\citenamefont {Dum}\ \emph {et~al.}(1992)\citenamefont {Dum},
  \citenamefont {Zoller},\ and\ \citenamefont {Ritsch}}]{Dum1992montecarlo}%
  \BibitemOpen
  \bibfield  {author} {\bibinfo {author} {\bibfnamefont {R.}~\bibnamefont
  {Dum}}, \bibinfo {author} {\bibfnamefont {P.}~\bibnamefont {Zoller}}, \ and\
  \bibinfo {author} {\bibfnamefont {H.}~\bibnamefont {Ritsch}},\ }\href
  {\doibase 10.1103/PhysRevA.45.4879} {\bibfield  {journal} {\bibinfo
  {journal} {Phys. Rev. A}\ }\textbf {\bibinfo {volume} {45}},\ \bibinfo
  {pages} {4879} (\bibinfo {year} {1992})}\BibitemShut {NoStop}%
\bibitem [{\citenamefont {M{\o}lmer}\ \emph {et~al.}(1993)\citenamefont
  {M{\o}lmer}, \citenamefont {Castin},\ and\ \citenamefont
  {Dalibard}}]{Molmer1993MonteCarlo}%
  \BibitemOpen
  \bibfield  {author} {\bibinfo {author} {\bibfnamefont {K.}~\bibnamefont
  {M{\o}lmer}}, \bibinfo {author} {\bibfnamefont {Y.}~\bibnamefont {Castin}}, \
  and\ \bibinfo {author} {\bibfnamefont {J.}~\bibnamefont {Dalibard}},\ }\href
  {\doibase 10.1364/JOSAB.10.000524} {\bibfield  {journal} {\bibinfo  {journal}
  {J. Opt. Soc. Am. B}\ }\textbf {\bibinfo {volume} {10}},\ \bibinfo {pages}
  {524} (\bibinfo {year} {1993})}\BibitemShut {NoStop}%
\bibitem [{\citenamefont {Plenio}\ and\ \citenamefont
  {Knight}(1998)}]{plenio1998quantumjumps}%
  \BibitemOpen
  \bibfield  {author} {\bibinfo {author} {\bibfnamefont {M.~B.}\ \bibnamefont
  {Plenio}}\ and\ \bibinfo {author} {\bibfnamefont {P.~L.}\ \bibnamefont
  {Knight}},\ }\href {\doibase 10.1103/RevModPhys.70.101} {\bibfield  {journal}
  {\bibinfo  {journal} {Rev. Mod. Phys.}\ }\textbf {\bibinfo {volume} {70}},\
  \bibinfo {pages} {101} (\bibinfo {year} {1998})}\BibitemShut {NoStop}%
\bibitem [{\citenamefont {Kramida}\ \emph {et~al.}(2020)\citenamefont
  {Kramida}, \citenamefont {{Yu.~Ralchenko}}, \citenamefont {Reader},\ and\
  \citenamefont {{and NIST ASD Team}}}]{NIST_ASD}%
  \BibitemOpen
  \bibfield  {author} {\bibinfo {author} {\bibfnamefont {A.}~\bibnamefont
  {Kramida}}, \bibinfo {author} {\bibnamefont {{Yu.~Ralchenko}}}, \bibinfo
  {author} {\bibfnamefont {J.}~\bibnamefont {Reader}}, \ and\ \bibinfo {author}
  {\bibnamefont {{and NIST ASD Team}}},\ }\href {https://physics.nist.gov/asd}
  {}\bibinfo {howpublished} {{NIST Atomic Spectra Database (ver. 5.8),
  [Online]. Available: {\tt{https://physics.nist.gov/asd}} [2021, September 1].
  National Institute of Standards and Technology, Gaithersburg, MD.}} (\bibinfo
  {year} {2020})\BibitemShut {NoStop}%
\bibitem [{\citenamefont {Taichenachev}\ \emph {et~al.}(2006)\citenamefont
  {Taichenachev}, \citenamefont {Yudin}, \citenamefont {Oates}, \citenamefont
  {Hoyt}, \citenamefont {Barber},\ and\ \citenamefont
  {Hollberg}}]{taichenachev2006magnetic}%
  \BibitemOpen
  \bibfield  {author} {\bibinfo {author} {\bibfnamefont {A.}~\bibnamefont
  {Taichenachev}}, \bibinfo {author} {\bibfnamefont {V.}~\bibnamefont {Yudin}},
  \bibinfo {author} {\bibfnamefont {C.}~\bibnamefont {Oates}}, \bibinfo
  {author} {\bibfnamefont {C.}~\bibnamefont {Hoyt}}, \bibinfo {author}
  {\bibfnamefont {Z.}~\bibnamefont {Barber}}, \ and\ \bibinfo {author}
  {\bibfnamefont {L.}~\bibnamefont {Hollberg}},\ }\href
  {https://link.aps.org/doi/10.1103/PhysRevLett.96.083001} {\bibfield
  {journal} {\bibinfo  {journal} {Physical review letters}\ }\textbf {\bibinfo
  {volume} {96}},\ \bibinfo {pages} {083001} (\bibinfo {year}
  {2006})}\BibitemShut {NoStop}%
\bibitem [{\citenamefont {Barker}\ \emph {et~al.}(2016)\citenamefont {Barker},
  \citenamefont {Pisenti}, \citenamefont {Reschovsky},\ and\ \citenamefont
  {Campbell}}]{Barker2016double}%
  \BibitemOpen
  \bibfield  {author} {\bibinfo {author} {\bibfnamefont {D.~S.}\ \bibnamefont
  {Barker}}, \bibinfo {author} {\bibfnamefont {N.~C.}\ \bibnamefont {Pisenti}},
  \bibinfo {author} {\bibfnamefont {B.~J.}\ \bibnamefont {Reschovsky}}, \ and\
  \bibinfo {author} {\bibfnamefont {G.~K.}\ \bibnamefont {Campbell}},\ }\href
  {\doibase 10.1103/PhysRevA.93.053417} {\bibfield  {journal} {\bibinfo
  {journal} {Phys. Rev. A}\ }\textbf {\bibinfo {volume} {93}},\ \bibinfo
  {pages} {053417} (\bibinfo {year} {2016})}\BibitemShut {NoStop}%
\bibitem [{\citenamefont {Filin}\ and\ \citenamefont
  {Safronova}(2021)}]{Safronova21Private}%
  \BibitemOpen
  \bibfield  {author} {\bibinfo {author} {\bibfnamefont {D.}~\bibnamefont
  {Filin}}\ and\ \bibinfo {author} {\bibfnamefont {M.}~\bibnamefont
  {Safronova}},\ }\href@noop {} {\enquote {\bibinfo {title} {private
  communication},}\ } (\bibinfo {year} {2021})\BibitemShut {NoStop}%
\bibitem [{\citenamefont {Martin}(2013)}]{Martin13}%
  \BibitemOpen
  \bibfield  {author} {\bibinfo {author} {\bibfnamefont {M.~J.}\ \bibnamefont
  {Martin}},\ }\emph {\bibinfo {title} {Quantum Metrology and Many-Body
  Physics: Pushing the Frontier of the Optical Lattice Clock}},\ \href@noop {}
  {Ph.D. thesis},\ \bibinfo  {school} {University of Colorado} (\bibinfo {year}
  {2013})\BibitemShut {NoStop}%
\bibitem [{6le()}]{6level_eqs}%
  \BibitemOpen
  \href@noop {} {\bibinfo  {journal} {Program examples using the Julia package
  QuantumCumulants.jl, which analytically derive the system and correlation
  function equations for the six-level and two-level laser model are given in
  the supplementary material. Additionaly we show an example to calculate the
  two-level system parameters from the pumped six-level atom, using the Julia
  package QuantumOptics.jl. The equations in these examples are solved
  numerically for our standard parameters}\ }\BibitemShut {NoStop}%
\bibitem [{\citenamefont {Kr{\"a}mer}\ \emph {et~al.}(2018)\citenamefont
  {Kr{\"a}mer}, \citenamefont {Plankensteiner}, \citenamefont {Ostermann},\
  and\ \citenamefont {Ritsch}}]{kramer2018quantumoptics}%
  \BibitemOpen
\bibfield  {journal} {  }\bibfield  {author} {\bibinfo {author} {\bibfnamefont
  {S.}~\bibnamefont {Kr{\"a}mer}}, \bibinfo {author} {\bibfnamefont
  {D.}~\bibnamefont {Plankensteiner}}, \bibinfo {author} {\bibfnamefont
  {L.}~\bibnamefont {Ostermann}}, \ and\ \bibinfo {author} {\bibfnamefont
  {H.}~\bibnamefont {Ritsch}},\ }\href {\doibase 10.1016/j.cpc.2018.02.004}
  {\bibfield  {journal} {\bibinfo  {journal} {Comput. Phys. Commun.}\ }\textbf
  {\bibinfo {volume} {227}},\ \bibinfo {pages} {109} (\bibinfo {year}
  {2018})}\BibitemShut {NoStop}%
\bibitem [{\citenamefont {Hunter}(2007)}]{hunter2007matplotlib}%
  \BibitemOpen
  \bibfield  {author} {\bibinfo {author} {\bibfnamefont {J.~D.}\ \bibnamefont
  {Hunter}},\ }\href@noop {} {\bibfield  {journal} {\bibinfo  {journal}
  {Computing in science \& engineering}\ }\textbf {\bibinfo {volume} {9}},\
  \bibinfo {pages} {90} (\bibinfo {year} {2007})}\BibitemShut {NoStop}%
\end{thebibliography}%
\end{document}